\input ./style/arxiv-general.cfg
\documentclass[aoas,MSNbibl,nameyear,seceqn,dvips]{arximspdf}
\makeatletter
   \@ifpackageloaded{graphicx}{}{\usepackage{graphicx}}
\makeatother
\usepackage{multirow,dcolumn}

\doi{10.1214/15-AOAS833}
\volume{9}
\issue{3}
\pubyear{2015}
\firstpage{1484}
\lastpage{1509}
\docsubty{FLA}
\setattribute{copyright}{owner}{In the Public Domain}

\makeatletter
\newcolumntype{d}[1]{D{.}{.}{#1}}
\newcommand{\iid}{\stackrel{\mathrm{IID}}{\sim}}
\newcommand{\ind}{\stackrel{\mathrm{ind}}{\sim}}
\makeatother

\begin{document}
\begin{frontmatter}

\title{Inferring constructs of effective teaching from classroom observations:
An application of Bayesian exploratory factor analysis without restrictions\thanksref{T1}}
\runtitle{Constructs of effective teaching}
\thankstext{T1}{Supported in
part by the Bill and Melinda Gates Foundation (52048).}

\begin{aug}
\author[A]{\fnms{J.~R.}~\snm{Lockwood}\corref{}\thanksref{m1}\ead[label=e1]{jrlockwood@ets.org}},
\author[B]{\fnms{Terrance~D.}~\snm{Savitsky}\thanksref{m2}\ead[label=e2]{savitsky.terrance@bls.gov}}\\
\and
\author[A]{\fnms{Daniel~F.}~\snm{McCaffrey}\thanksref{m1}\ead[label=e3]{dmccaffrey@ets.org}}
\runauthor{J.~R. Lockwood, T.~D. Savitsky and D.~F. McCaffrey}

\affiliation{Educational Testing Service\thanksmark{m1} and U.S.
Bureau of Labor Statistics\thanksmark{m2}}
\address[A]{J.~R. Lockwood\\
D.~F. McCaffrey\\
Educational Testing Service\\
660 Rosedale Road\\
Princeton, New Jersey 08541\\
USA\\
\printead{e1}\\
\phantom{E-mail:\ }\printead*{e3}
}

\address[B]{T.~D. Savitsky\\
U.S. Bureau of Labor Statistics\\
2 Massachusetts Ave. N.E\\
Washington, DC 20212 \\
USA\\
\printead{e2}}
\end{aug}

%
\received{\smonth{6} \syear{2014}}
%
\revised{\smonth{3} \syear{2015}}

%
\begin{abstract}
Ratings of teachers' instructional practices using standardized classroom
observation instruments are increasingly being used for both research and
teacher accountability. There are multiple instruments in use, each attempting
to evaluate many dimensions of teaching and classroom activities, and little
is known about what underlying teaching quality attributes are being
measured. We use data from multiple instruments collected from 458 middle
school mathematics and English language arts teachers to inform
research and
practice on teacher performance measurement by modeling latent
constructs of
high-quality teaching. We make inferences about these constructs using a
novel approach to Bayesian exploratory factor analysis (EFA) that, unlike
commonly used approaches for identifying factor loadings in Bayesian
EFA, is
invariant to how the data dimensions are ordered. Applying this
approach to
ratings of lessons reveals two distinct teaching constructs in both
mathematics and English language arts: (1) quality of instructional practices;
and (2) quality of teacher management of classrooms. We demonstrate the
relationships of these constructs to other indicators of teaching quality,
including teacher content knowledge and student performance on standardized
tests.
\end{abstract}


\end{frontmatter}

\section{Introduction}\label{sec1}
National, state and local education policy is undergoing a dramatic shift
focused on individual teacher accountability. Encouraged by federal initiatives
such as the Race to the Top grant competition, state legislation
mandating that
teacher evaluations based on individual performance measures be used for
consequential decisions such as pay or retention is rapidly diffusing
across the
nation. Numerous instruments for measuring the quality of teaching are being
used or developed, including measures of instructional practices, teacher
subject-matter and pedagogical knowledge, quality and rigor of work
assigned to
students, student perceptions of teacher quality, and student learning outcomes
[\citet{MET:2013}]. While there is general agreement that these
measures are
important, it is not well understood what underlying constructs define
``teaching quality'' and to what extent different measures capture these
constructs. We do know that the quality of teachers' instructional
practice is
modest for the majority of teachers in research studies
[\citet{gitomer:etal:2014,MET:2013}]. We also know that student
achievement in
the United States lags behind other countries and falls short of our own
national standards [\citet{peterson2011globally}]. The goal of restructuring
teacher evaluation systems is to change these circumstances by
improving the
average quality of teaching in the teacher workforce.

Yet, without understanding the underlying constructs that define teaching
quality, it is difficult to design systems to achieve this goal. If the
constructs that define high-quality teaching are not easily malleable,
the most
effective systems might focus on hiring strong teachers and firing weak teachers
[\citet{Gord:Kane:Stai:2006}]; however, if the constructs are not
intrinsic to
individuals, then systems might instead focus on improving teaching practice
through professional development. Therefore, both what constructs to measure
and how to use those measures to take action require understanding what makes
an effective teacher capable of promoting student learning.

We contribute to this goal by investigating the underlying constructs of
high-quality teaching using data from over 450 middle school teachers who
participated in the Understanding Teacher Quality (UTQ) study
(\surl{www.utqstudy.org}). The data include ratings of participating teachers'
instructional practices from four different standardized instruments
that were
developed from different theoretical perspectives on teaching quality. Our
primary research question is whether those perspectives are defining
common or
distinct teaching quality constructs, which we address using
exploratory factor
analysis (EFA) on the instructional practice ratings to uncover latent teaching
quality attributes. We perform the factor analysis within a latent hierarchical
model for the ordinal instructional ratings to separate the teacher-level
variation, of direct interest, from the other sources of variance such as
day-to-day lesson variation and errors introduced by the raters who assign
scores. We develop a novel Bayesian implementation of this model that improves
upon existing Bayesian approaches for EFA. We then examine how
estimated factor
scores extracted from the instructional practice ratings relate to assessments
of teacher knowledge and teacher impacts on student achievement growth to
provide validity evidence about the latent constructs. Collectively, our
investigations provide an important step toward validating commonly used
measures as providing useful indicators of teaching quality, and offer insight
into the distinguishable components of teaching.

\section{Understanding teaching quality data}

The UTQ study took place in middle schools of three large school
systems from
the same United States metropolitan region. It includes 458 teachers teaching
mathematics ($n=231$) or English language arts (ELA; $n=227$) to 6th--8th graders
(typically ages 11--14). Participation in the study was voluntary. Data were
collected over two years, with about half of the teachers participating
in each
year.

From each participating teacher we collected three types of measures: (1)
evaluations of instruction based on ratings of video-recorded lessons,
(2) scores
on a teacher knowledge test, and (3) estimates of teachers' effects on student
standardized achievement tests. In this section we describe the
evaluations of
instruction based on ratings of video-recorded lessons. We describe the other
two measures in Section~\ref{ssec:stage2} where we examine their relationships
to the constructs derived from the lesson ratings.

For each study teacher, four lessons were video recorded during the
school year.
The study schools followed a traditional middle school format where
each teacher
taught multiple classrooms across different periods of the day. For each
teacher we sampled two study classrooms, which we refer to as the two different
\emph{sections} for that teacher, and for each section we recorded two lessons
from different days. For the purposes of applying the rating
instruments, a
lesson is divided into a set of disjoint time intervals called \emph{segments}
lasting seven, 15, 30 or 45 minutes, depending on the rating instrument.

%
\begin{table}[b]
\caption{Summary of protocols used to rate instructional practice}
\label{tab:protocols}
\begin{tabular*}{\textwidth}{@{\extracolsep{\fill}}llcc@{}}
\hline
\textbf{Instrument} & \multicolumn{1}{c}{\textbf{Description}} & \textbf{\# Dimensions} & \textbf{Scale} \\
\hline
CLASS &Classroom Assessment \& Scoring System & $10$ & 1--7\\
FFT &Framework for Teaching & $11$ & 1--4 \\
PLATO &Protocol for Language Arts Teaching & $13$ & 1--4\\
MQI &Mathematics Quality of Instruction & \phantom{0}$8$ & 1--3\\
\hline
\end{tabular*}
\end{table}

Video-recorded lessons were rated using four different standardized observation
instruments (or ``protocols''), summarized in Table~\ref
{tab:protocols}. Each
instrument consists of multiple dimensions. The \emph{Classroom
Assessment and
Scoring System} [CLASS; \citet{hamre:2012}] measures $10$ dimensions of
classroom interactions including the teachers' management and
organization of
the classroom, their engagement of and responsiveness to students, and aspects
of their instruction. The \emph{Framework for Teaching} [FFT; \citet{fft}]
consists of $11$ dimensions focusing on the domains of classroom
environment and
quality of instruction. The \emph{Protocol for Language Arts Teaching
Observations} [PLATO; \citet{grossman:2010}] is specific to ELA and defines
$13$ dimensions that measure specific instructional practices,
strategies for
encouraging student participation, behavioral management and time management.
Finally, the \emph{Mathematical Quality of Instruction} [MQI; \citet{mqi}]
evaluates various aspects of mathematics instruction; for this study we
focus on
$8$ of these dimensions. Two of the instruments (CLASS and FFT) apply
to both
math and ELA instruction, while the others (PLATO for ELA and MQI for
math) are
specific to only one subject. All four instruments use ordered scores intended
to record the level of quality expressed in each dimension. Further
details on
the dimensions are provided in Table~\ref{tab:loadings} in the
\hyperref[app]{Appendix}.

Eleven raters conducted all scoring of the video-recorded lessons, six
with math
expertise and five with ELA expertise. All raters scored using CLASS and
FFT. Only raters with the corresponding subject expertise scored using
MQI and
PLATO. Raters received extensive training in all instruments and demonstrated
proficiency prior to rating lessons. They also underwent regular calibration
checks for the duration of scoring to promote accuracy in scores. See
\citet{casa:lock:mcca:2015} for details.

The lesson scoring data are multivariate with a combination of nested and
crossed structures. There are $458$ teachers, $916$ sections (two for each
teacher), $1828$ video-recorded lessons (two for each section except
for a tiny
amount of missing data) and $6141$ segments (approximately 3--4 per
lesson). These units are structured hierarchically. Each lesson was
scored on
exactly three instruments: CLASS, FFT, and one of PLATO or MQI. A
scoring event
consists of a rater assigning a vector of scores to the dimensions of a
particular instrument for each segment of the lesson. For each instrument,
about 80\% of the lessons were scored by a single rater, while the remainder
were scored by two separate raters. The rating process introduces partial
crossing because for each instrument, each rater scored lessons from multiple
different teachers and sections, but all raters do not score lessons
from all
teachers on any instrument, and no lessons were scored by all raters.

Our goal was to test if teaching quality observed in classrooms can be
decomposed into a lower-dimensional set of latent teaching quality
constructs. We used the ratings data on all dimensions of the observation
instruments ($34$ dimensions across three instruments for ELA, and $29$
dimensions across three instruments for math) to conduct EFA at the teacher
level.
The measurement structure for the instructional practice ratings is
complex when
viewing the scores as indicators of constructs for individual teachers:
we have
multivariate ordinal categorical data from multiple instruments, and
all scores
are contaminated by errors related to the particular sections, lessons, and
raters who scored the lesson, with errors at all levels potentially being
correlated across dimensions. As demonstrated by \citet
{mcca:class:2015}, not
accounting for these errors can distort inferences about factor
structure at the
teacher level. Likelihood approaches to estimating factor structure at the
teacher level would be challenged by the large number of dimensions,
the ordinal
data, and the mixed hierarchical and crossed measurement structure. Bayesian
approaches simplify the estimation of a model requiring integration
over so many
latent variables where both the teacher factor structure and aspects of the
measurement process are modeled. We thus proceed in Section~\ref
{hiermod} by
presenting a hierarchical model for the ratings which includes a standard
exploratory factor model at the teacher level. We then present a method for
conducting Bayesian EFA to yield interpretable factors to support our
goal of
understanding the constructs of teaching, starting with a discussion of a
practical problem with Bayesian EFA in Section~\ref{befa}, then
turning to our
solution to that problem in Section~\ref{pibefa}. We present results
of our
application in Section~\ref{case} and concluding remarks in
Section~\ref{discussion}.

\section{Model for instructional ratings data}\label{hiermod}

\subsection{Relating observations to latent effects}

We model the data from each subject (math and ELA) separately. For each subject,
the data consist of vectors of scores from $N$ scoring events. For a scoring
event, a rater, using one of the three instruments, assigned scores on
all the
dimensions of the instrument for a segment of a lesson taught by one of the
study teachers to one of two of the study sections for that teacher. We index
such observations by $i$. For each subject, the data have $j = 1,
\ldots,N_{\mathrm{teach}}$ teachers and we use $j_i$ to identify the teacher
whose lesson was scored in observation $i$. Similarly, there are $s =
1, \ldots,
N_{\mathrm{sect}}$ sections and $v=1, \ldots, N_{\mathrm{lesson}}$
lessons, and
we use $s_i$ and $v_i$ to denote the section and lesson corresponding to
observation $i$. Finally, there are $r=1, \ldots, N_{\mathrm{rater}}$
raters for
each subject and $r_i$ denotes the rater who conducted observation $i$.
We let
$\mathcal{P}_i$ denote the instrument (protocol) used for scoring observation
$i$. For math, $\mathcal{P}_i \in\{\mathrm{CLASS},\mathrm{FFT},\mathrm
{MQI}\}$ and
for ELA, $\mathcal{P}_i \in\{\mathrm{CLASS},\mathrm{FFT},\mathrm{PLATO}\}
$. We let
$\mathbf{y}_i$ denote the vector of scores assigned by the rater for observation
$i$ and
$y_{id}$ be the score on dimension $d$, $d = 1, \ldots, D_{\mathcal
{P}_i}$. Each
$y_{id}$ takes one of a discrete set of possible ordinal scores that
depends on
the protocol,
$y_{id} \in\{1, \ldots, L_{\mathcal{P}_i}\}$.


We assume that each ordinal score $y_{id}$ has a latent $t_{id}$ such that
\begin{eqnarray*}
y_{id} = \ell\in\{1,\ldots,L_{\mathcal{P}_i}\} &\Leftrightarrow&
\gamma_{\mathcal{P}_i, d,\ell- 1} < t_{id} \leq\gamma_{\mathcal
{P}_i, d,\ell},
\\
t_{id} \vert\mu_{id} &\ind& \mathcal{N} (
\mu_{id}, 1 ),
\end{eqnarray*}
as described in \citet{Albe:Chib:1993}, \citet{congdon:2005},
\citet{Johnson:1996} and \citet{savitsky:mccaffrey:2014}. We model
$\bolds{\mu}_i=(\mu_{i1}, \ldots, \mu_{iD_{\mathcal{P}_i}})$ as
%
%
\begin{equation}
\label{eq:mui} \bolds{\mu}_i = \bolds{\delta}_{j_i,\mathcal{P}_i} +
\bolds{\phi}_{s_i,\mathcal{P}_i} + \bolds{\theta}_{v_i,\mathcal{P}_i} +
\bolds{\kappa}_{r_i,\mathcal{P}_i} +
\bolds{\zeta}_{v_i, r_i,\mathcal{P}_i},
\end{equation}
where $\bolds{\delta}_{j_i,\mathcal{P}_i} =$ the vector of teacher
effects for teacher $j_i$;
$\bolds{\phi}_{s_i,\mathcal{P}_i} =$ the vector of section effects
for section $s_i$;
$\bolds{\theta}_{v_i,\mathcal{P}_i} =$ the vector of lesson effects for
lesson $v_i$;
$\bolds{\kappa}_{r_i,\mathcal{P}_i} = $ the vector of rater effects for rater
$r_i$; and
$\bolds{\zeta}_{v_i, r_i,\mathcal{P}_i} =$ the vector of rater by lesson
effects for
lesson $v_i$ and rater $r_i$. Each is a vector of $D_{\mathcal{P}_i}$ effects
for the dimensions of protocol $\mathcal{P}_i$.

The model for $\bolds{\mu}_i$ does not include terms for either
segments or
rater by segment interactions. Hence, any variability in scores due to those
sources is captured by $\operatorname{Var}(t_{id} \vert\mu_{id})$, which is specified
as 1. In addition, any nonzero covariances in rater errors in the dimension
scores for a segment, like those found by \citet{mcca:class:2015}, will
contribute to the covariances among the elements of the rater by lesson effects,
$\bolds{\zeta}_{v_i, r_i,\mathcal{P}_i}$.

Our goal is to study the structure among the dimensions from all the protocols
used in each subject. Hence, we need to jointly model the random
effects from
all the protocols. To do this for math teachers, we define for each teacher
$j=1,\ldots,N_{\mathrm{teach}}$ the combined vector of teacher effects
$\bolds{\delta}_{j} =
(\bolds{\delta}_{j,\mathrm{CLASS}}',\bolds{\delta}_{j,\mathrm{FFT}}',\bolds{\delta}_{j,\mathrm{MQI}}')'$
with elements $\delta_{jq}$ for $q=1, \ldots, D_{\mathrm{math}}$, where
$D_{\mathrm{math}}= D_{\mathrm{CLASS}} + D_{\mathrm{FFT}} + D_{\mathrm{MQI}} = 29$, the total number of dimensions across the three
protocols. We use the subscript $j$ rather than $j_i$ because we are referring
to the effects for teacher $j$ that apply to all of the observations
$i$ for
which he or she is the corresponding teacher. We similarly define
$\bolds{\phi}_{s}$ and $\bolds{\theta}_{v}$ for the classes and
lessons, and
$\bolds{\kappa}_{r}$ for the raters. The rater by lesson interactions are
protocol-specific because any given rater uses only one protocol to
score any
given lesson. Hence, we do not use combined vectors for these effects.
We define
the analogous set of combined teacher, section, lesson, and rater
random effect
vectors for the ELA data. These vectors have $D_{\mathrm{ELA}}=34$ elements
corresponding to the total number of dimensions in the three protocols
used to
score ELA observations.

\subsection{Model for the latent effects}

To complete the model, we need to specify priors for the cutpoints that link
the
ordinal observed scores to the latent variables, and priors for the random
effects. For a given dimension $d$ of a protocol $\mathcal{P}$, we define
$\gamma_{\mathcal{P}, d, 0} = -\infty$ and $\gamma_{\mathcal{P}, d,
L_{\mathcal{P}}} = \infty$, but must specify priors for the remaining
$L_{\mathcal{P}} - 1$ cutpoints. These cutpoints can be estimated from
the data
because (1)~we fixed the conditional variance of $t_{id}$ to be 1; (2)~multiple
scores given by an individual rater to segments from the same lesson
share a
common $\mu_{id}$; and (3)~the marginal mean of $\mu_{id}=0$ since,
as discussed
below, each of the latent effects in equation~(\ref{eq:mui}) is mean
zero. To
specify the prior for unknown cutpoints, we follow \citet
{Ishw:univ:2000} and
assume $\gamma_{d,\ell} \equiv
\mathop{\sum}_{l=1}^{\ell}\exp(\rho_{d,l} )$, where $\rho_{d,l}
\sim
\mathcal{N} (0,\tau_{d}^{2} )$ and $\tau_{d} \iid
\operatorname{Uniform}(0,100)$, without order restrictions. We selected this
prior as a
possible means of improving mixing on draws for the cutpoints
[\citet{savitsky:mccaffrey:2014}].

For teacher effects, we specify a factor model for the $D \times1$ vectors
$\{\bolds{\delta}_{j}\}$ of combined effects from all three protocols for
teachers in each subject area:
%
%
\begin{equation}
\label{eq:efa} \bolds{\delta}_{j} = \bolds{\Lambda}\bolds{
\eta}_{j} + \bolds{\varepsilon}_{j}.
\end{equation}
Here $\bolds{\Lambda}$ is the $D \times K$ loadings matrix and
$\bolds{\eta}_{j}$
is the $K \times1$ vector of factor scores for teacher $j$, where $K$ denotes
the number of factors. We drop the subject-specific subscript in $D$ to simplify
the presentation, but the dimensions will differ for math and ELA. The
uniqueness is $\bolds{\varepsilon}_{j} \iid
\mathcal{N}_{D} (\mathbf{0},\mathbf{U} )$, where $\mathbf{U}$
is the diagonal matrix of uniqueness variances. We specify $\bolds
{\eta}_{j}
\sim\mathcal{N}_{K} (\mathbf{0},\mathbf{I}_{K} )$ to identify the
scale of loadings. Marginalizing over the factors gives
$\operatorname{Cov} (\bolds{\delta}_{j} ) =
\bolds{\Lambda}\bolds{\Lambda}^{\prime} + \mathbf{U} = \mathbf
{Q} + \mathbf{U}$,
with communality, $\mathbf{Q}$, and uniqueness, $\mathbf{U}$. Additional
information about our prior distributions for the loadings and uniqueness
variances are in Section~\ref{iidprior}. We model the remaining random effects
from equation~(\ref{eq:mui}) as multivariate Gaussian with mean zero
and a
precision matrix that has a Wishart prior with an identity scale matrix and
degrees of freedom equal to one plus the dimension of the random effect vectors.

\subsection{Identification issues in EFA}\label{ident}

A well-known limitation of the factor model~(\ref{eq:efa}) is that
there is no
unique set of loadings. Orthogonal rotations of the loadings and factor scores
yield identical values of $\bolds{\delta}$. For any $K \times K$ orthogonal
rotation matrix $\mathbf{P}'$, if $\bolds{\Lambda}^* =
\bolds{\Lambda}\mathbf{P}'$ and $\bolds{\eta}^*=\mathbf{P}\bolds
{\eta}$, then
$\bolds{\Lambda}^*\bolds{\eta}^* =
\bolds{\Lambda}\mathbf{P}'\mathbf{P}\bolds{\eta} =
\bolds{\Lambda}\bolds{\eta}$. The loadings are not identified by the
likelihood; rather, the communality matrix $\mathbf{Q}$ is identified.
That is,
for any $D \times K$ full-column rank loadings matrices, $\bolds
{\Lambda}$ and
$\bolds{\Lambda}^*$ where $\bolds{\Lambda}^* = \bolds{\Lambda
}\mathbf{P}'$
for some $K \times K$ orthogonal rotation matrix, $\mathbf{Q}^* =
\bolds{\Lambda}^*\bolds{\Lambda}^{*'}$ is equal to $\mathbf{Q} =
\bolds{\Lambda}\bolds{\Lambda}'$. In maximum likelihood (MLE)
inference, the
lack of identification of the loadings is resolved by picking an arbitrary
$\bolds{\Lambda}$ such that $\bolds{\Lambda} \bolds{\Lambda}' =
\widehat{\mathbf{Q}}_{\mathrm{MLE}}$ and then rotating $\bolds{\Lambda}$ to meet
criteria for interpretability. A common goal is to seek a rotation that results
in a so-called ``simple structure'' of the loadings where each
dimension loads
relatively strongly on one factor and weakly on all others. Simple
structure is
encouraged by choosing loadings that optimize an external criterion
such as
varimax [\citet{kaiser:1958}] or related criteria [\citet
{browne:2001}]. However,
we want to conduct a Bayesian analysis and determine if a simple interpretable
factor structure exits. Bayesian methods to identify the factors use different
criteria, so we must modify the traditional methods, which we now describe.

\section{Bayesian EFA}\label{befa}

Bayesian EFA models commonly identify loadings separately from factors by
restricting the structure of the loadings matrix to be lower
triangular, with
nonnegative diagonals to account for sign reflections, and then specifying
priors for the free parameters of the resulting constrained loadings matrix
[\citet{Gewe:Zhou:meas:1996,Lope:West:baye:2004}].\footnote{Note lower
triangular
structure is not required for identification. Identification requires elements
of the columns of the loadings matrix to be zero but the ordering of those
columns does not matter.} This restriction yields a unique loadings
representation
[\citet{fsl:2013}]. The row index of each leading nonzero factor loading
increases from left to right along the diagonal under the lower triangular
restriction. The dimension associated with a leading nonzero loading
for a
factor is referred to as a ``founder'' dimension for that factor
[\citet{Carv:Chan:Luca:Nevi:Wang:West:high:2008}].

This approach has a few disadvantages for our application. First, the
restriction to lower triangular loadings matrices is not substantively
motivated. This restriction is chosen solely for identification. In other
applications, lower triangular loadings may support a substantive interpretation
and these constraints may be appropriate; see, for example,
\citet{hahn:carv:scott:2012}. However, that is not the case with teacher
observations.


Second, the lower triangular restriction induces a prior for the communality
$\mathbf{Q}$ that is sensitive to the ordering of the dimensions
[\citet
{Bhat:Duns:spar:2011,Carv:Chan:Luca:Nevi:Wang:West:high:2008,fsl:2013,mcpa:ETAL:2014}].
Specifically,
assuming exchangeable prior distributions for nonzero loadings under
the lower
triangular restriction, the induced prior distributions for elements of
$\mathbf{Q}$ associated with founder dimensions
[\citet{Carv:Chan:Luca:Nevi:Wang:West:high:2008}] are different than
those for
elements of $\mathbf{Q}$ associated with other dimensions.
Thus, for given matrices $\mathbf{Q}$ and $\mathbf{Q}^*$ where
$\mathbf{Q}^*$
equals $\mathbf{Q}$ with its row and column elements permuted as they
would be
if we permuted the order of the variables, the induced prior
probability on
$\mathbf{Q}$ does not equal the induced prior probability on $\mathbf{Q}^*$.
Our inferences about communalities, and consequently about any rotation
of the
loadings, would be sensitive to variable ordering. This is unlike the
MLE EFA
solution, where the permutation invariance of the likelihood function implies
that a permutation of $\widehat{\mathbf{Q}}_{\mathrm{MLE}}$ is equal to the
MLE solution
$\widehat{\mathbf{Q}}^{*}_{\mathrm{MLE}}$ under the permuted data, and so
inferences with
respect to any optimized rotation criterion that does not depend on variable
ordering will also be permutation invariant.

The sensitivity to variable ordering is potentially problematic in our
application. We are interested in factor structure at the teacher
level, which
must be inferred with only about 225 teachers per subject using coarsened
ordinal data subject to multiple sources of nuisance measurement error
(e.g., sections, lessons, segments and raters). The amount of data information
about the constructs of interest may not overwhelm the prior distribution,
leaving us potentially vulnerable to sensitivities to variable ordering imposed
by the prior. Also, the computational burdens of estimating the model in
Section~\ref{hiermod} precludes trying many different orderings of the
variables
to explore sensitivity of the findings. Thus, our goal was to use a prior
distribution that is exchangeable across dimensions so that the prior
probability on any communality matrix $\mathbf{Q}$ equals the prior probability
on $\mathbf{P}\mathbf{Q}\mathbf{P}'$, where $\mathbf{P}$ is a $(D
\times D)$
permutation matrix. When combined with an exchangeable prior
distribution for
the uniqueness variances $\mathbf{U}$, this would provide Bayesian EFA
inferences that shared the same permutation invariance as MLE EFA.


\subsection{Alternative Bayesian identification strategies} \label{altmeth}


An alternative to sampling loadings is to sample the communality and derive
loadings from it. The communality is identified and, moreover, every
$\mathbf{Q}$ defines a unique infinite set of loadings matrices
$\bolds{\Lambda}$, such that $\bolds{\Lambda}\bolds{\Lambda
}'=\mathbf{Q}$.
Hence, if a satisfactory prior for the communality can be specified, inferences
about loadings can be made by setting a rule to select a loading matrix
from the
set of loadings associated with the communality. However, because the
communality is not full rank, standard conjugate or other widely used
priors for
random positive definite symmetric matrices cannot be used. \citet
{carmeci:2008}
directly samples the rank-deficient $\mathbf{Q}$ through a Metropolis--Hastings
scheme with a prior distribution specified as a mixture of singular Wishart
distributions. He pointed out that his approach is computationally burdensome
compared to directly sampling the loadings matrix, such that it is recommended
only for small and medium size factor models. Given we have $34$
dimensions for
ELA and $29$ for math and we are conducting EFA in the context of
a cross-classified, hierarchical, ordinal data model, which also increases
computational time, this solution was unacceptable for our case study. His
approach also requires a specialized MCMC sampler, and we were
interested in an
approach that could be straightforwardly coded in the BUGS language.


\citet{Carv:Chan:Luca:Nevi:Wang:West:high:2008} use the lower triangular
restriction and incorporate selection of founders into their model to find
dimensions with high probabilities for having nonzero founder loadings, though
they did not address nonexchangeability of the induced priors for the
communality parameters among the dimensions. \citet{fsl:2013}
addressed the
prior sensitivity to dimension ordering by making inferences about a generalized
lower triangular matrix, which is a matrix in which all the elements
above the
diagonal are zero but some of the diagonal and lower triangular
elements can be
zero. As with the lower triangular matrix, we did not have a specific
substantive interest in loadings from the generalized lower triangular
matrix. \citet{fsl:2013} state that their method ``handles the
ordering problem
in a more flexible way'' (page 4), but they do not specifically address
the issue of
exchangeability of the induced prior on the communalities. Moreover,
even if
their approach induces an exchangeable prior, their method requires a
specialized MCMC sampler.

\citet{Bhat:Duns:spar:2011} introduce a class of shrinkage priors
intended to
estimate reduced-rank covariance matrices for high-dimensional data.
This can
be used to obtain a permutation-invariant prior distribution for
$\mathbf{Q}$,
but by construction will tend to shrink away weakly expressed factors.
In our
application we anticipated that factors could be weakly expressed
because of
both the possible subtleties inherent to effective teaching and the
fact that
our measures on teachers are contaminated by relatively large
measurement errors
at the section, lesson and rating level. We thus determined this
approach would
not be suitable for our application. Rather, we blend the ideas of
\citet{Bhat:Duns:spar:2011} of obtaining a permutation-invariant prior
distribution for $\mathbf{Q}$ with the parameter-expansion approach to
parameterizing loadings of \citet{Ghos:Duns:defa:2009} to induce a prior
distribution for $\mathbf{Q}$ that is better tuned to our application.
We next
describe our prior specification and our procedure for determining identified
loadings.


\section{Permutation-invariant Bayesian EFA}\label{pibefa}

We use a three-step approach to sample communalities and derive our final
loadings estimates in a manner that yields permutation-invariant inferences
about loadings for the factor structure. In the first step we model the
elements of an unrestricted $\bolds{\Lambda}$ with exchangeable prior
distributions to induce a prior distribution on the communality
$\mathbf{Q}$
that is permutation invariant. When combined with an exchangeable prior
for the
uniqueness variances $\mathbf{U}$, this achieves the goal of having a
permutation-invariant prior distribution for
$\operatorname{Cov} (\bolds{\delta}_{j} ) = \mathbf{Q} + \mathbf{U}$. In the
second step, we rotate sampled $\bolds{\Lambda}$ to obtain loadings
with simple
structure using the varimax criterion [\citet{kaiser:1958}]. Finally, because
loadings meeting the varimax criterion are not unique ($2^KK!$
solutions exist
by permuting or changing the signs of columns of any given solution),
the third
step of our approach reorients the varimax rotations draw by draw to
move them
all to a common orientation. We describe each of these steps in turn.

\subsection{Exchangeable priors on loadings and uniqueness} \label{iidprior}

The key requirements of our approach are (1) to place no restrictions
on the
elements $\lambda_{dk}$ of the working loadings matrices $\bolds
{\Lambda}$
(e.g., do not use lower triangular restrictions); and (2)~to use
exchangeable prior
distributions for the $\lambda_{dk}$. These two conditions ensure that if
$\mathcal{G}_{[ij]}(q)$ is the induced prior for the row $i$ and
column $j$
element of $\mathbf{Q}$, then $\mathcal{G}_{[ii]}(q) =\mathcal
{G}_{[i'i']}(q)$
for any $i$ and $i'$ and $\mathcal{G}_{[ij]}(q) =\mathcal
{G}_{[i'j']}(q)$ for
any $i$, $j$, $i'$, $j'$ where both $i \neq j$ and $i' \neq j'$. That
is, there
is one common exchangeable prior for the diagonal elements of $\mathbf
{Q}$ and
another common exchangeable prior for the off-diagonal elements. This
makes the
induced prior for $\mathbf{Q}$ invariant to permutations of the data
dimensions.

Any exchangeable prior distribution for $\lambda_{dk}$ would suffice, including
IID, but we adopt the parameter expansion approach of
\citet{Ghos:Duns:defa:2009} to improve mixing of the working loadings.
We use
the following reparameterized model:
\begin{eqnarray*}
\bolds{\delta}_{j} &=& \bolds{\Lambda}^{\#}\bolds{
\eta}^{\#}_{j} + \bolds{\varepsilon}_{j},
\\
\bolds{\eta}^{\#}_{j} &\iid& \mathcal{N} \bigl(\mathbf{0},
\bolds{\Phi}^{-1} \bigr),
\\
\bolds{\Phi} &=& \operatorname{diag}(\phi_{1},\ldots,\phi_{K}),
\end{eqnarray*}
where the elements $\lambda^{\#}_{dk}$ of $\bolds{\Lambda}^{\#}$ are modeled
with independent standard normal priors and $\phi_{k}^{-1}$ are IID
$\operatorname{Gamma}(a,b)$ with common mean $a/b$ and variance $a/b^2$. We use
$a=b=1.5$. The inverse transforms $\lambda_{dk} =
\lambda^{\#}_{dk}\phi_{k}^{-{1}/{2}}$ and $\eta_{jk} =
\eta^{\#}_{jk}\phi_{k}^{{1}/{2}}$ remove the redundant $\bolds
{\Phi}$ and
induce a marginal $t$ prior for $\lambda_{dk}$.

To complete the permutation invariance of the prior distribution for
the factor
model, we also need an exchangeable prior on the diagonal elements of
$\mathbf{U}$, $u_{dd}$, $d=1, \ldots, D$. Following the common approach,
$u_{dd}^{-1}$ are IID\break $\operatorname{Gamma}(a,b)$ with $a=b=1.5$. Again, any
exchangeable
prior would suffice. We also tested sensitivity to an alternative prior
distribution where the square roots of the $u_{dd}$ were modeled as IID uniform
[\citet{Gelm:2006}]. Inferences about the latent teaching constructs
and their
relationships to other teaching quality indicators were not sensitive
to this
alternative prior.

\subsection{The varimax rotation}\label{ssec:varimax}

In the second step, for each $\bolds{\Lambda}_{b}$, $b=1, \ldots, B$ sampled
from the posterior where $B$ is the total number of MCMC samples, we rotate
$\bolds{\Lambda}_{b}$ to obtain loadings satisfying the varimax criterion
[\citet{kaiser:1958}]. Specifically, given a candidate loadings matrix
$\bolds{\Lambda}$, the varimax criterion results in loadings
$\bolds{\Lambda}\mathbf{R}_V(\bolds{\Lambda})$ where
\[
\mathbf{R}_V(\bolds{\Lambda}) = \mathop{\operatorname{arg\,max}}_{\mathbf{R}} \sum_{k=1}^{K} \Biggl(
\frac{1}{D}\sum_{d=1}^{D} (\bolds{
\Lambda}\mathbf{R})_{dk}^4 - \Biggl( \frac{1}{D} \sum
_{d=1}^{D}(\bolds{\Lambda}
\mathbf{R})_{dk}^2 \Biggr)^2 \Biggr),
\]
and $(\bolds{\Lambda}\mathbf{R})_{dk}$ denotes the $d,k$ element of
the matrix
$\bolds{\Lambda}\mathbf{R}$. The notation $\mathbf{R}_V(\bolds
{\Lambda})$ is
used to emphasize that the chosen rotation matrix depends on the input matrix
$\bolds{\Lambda}$. However, the final varimax loadings
$\bolds{\Lambda}\mathbf{R}_V(\bolds{\Lambda})$ are specific to the
communality
matrix $\mathbf{Q}$ in that if $\bolds{\Lambda}$ and $\bolds
{\Lambda}^{*}$
satisfy $\bolds{\Lambda}\bolds{\Lambda}' =
\bolds{\Lambda}^*\bolds{\Lambda}^{*'} = \mathbf{Q}$, then
$\bolds{\Lambda}\mathbf{R}_V(\bolds{\Lambda}) =
\bolds{\Lambda}^{*}\mathbf{R}_V(\bolds{\Lambda}^{*})$ up to an equivalence
class of $2^KK!$ matrices that differ by $2^K$ column sign reflections
and $K!$
column permutations. That is, for a given $\mathbf{Q}$ there are $2^KK!$
loadings matrices that meet the varimax criterion, differing only by column
order and sign. For each draw we obtain $\mathbf{R}_V(\bolds{\Lambda
}_{b})$ and
$\bolds{\Lambda}_{Vb} =
\bolds{\Lambda}_{b}\mathbf{R}_V(\bolds{\Lambda}_{b})$. However, we cannot
guarantee that all draws are oriented to the same column ordering and
sign. Hence, by using the varimax criterion to select loadings for interpretable
factors, we reduced the infinite dimensional problem of selecting a loadings
matrix from $\mathbf{Q}$ to a $2^KK!$ dimensional problem of selecting the
orientation of varimax solutions.

\subsection{Identifying varimax loadings}\label{ssec:loadings}

In our final step we reorient the varimax loadings from each draw,
$\bolds{\Lambda}_{Vb}$, to a common orientation. The need for post hoc
reorientation of samples to deal with indeterminacies in Bayesian factor
analysis is commonplace, and our approach is similar to ones developed by
\citet{hoff:raft:hand:2002}, \citet{fsl:2013}, \citet
{erosheva:curtis:2013} and
\citet{mcpa:ETAL:2014}, as well as that of \citet{stephens:2000} for mixture
models.

Following \citet{hoff:raft:hand:2002} and \citet{mcpa:ETAL:2014}, we
select the
orientation $\bolds{\Lambda}_{Vb}$ which makes each of its columns
closest, in
Euclidean distance, to the columns of a reference matrix. That is, for
a given
target $\bolds{\Lambda}_{Vb^{*}}$ we find the matrix $\mathbf{T}_b$ that
minimizes
%
%
\begin{equation}
\label{eq:orient} \operatorname{tr} \bigl[(\bolds{\Lambda}_{Vb^{*}}-\bolds{
\Lambda}_{Vb}\mathbf{T}_b)' (\bolds{
\Lambda}_{Vb^{*}}-\bolds{\Lambda}_{Vb}\mathbf{T}_b)
\bigr]
\end{equation}
among all of the $2^K K!$ matrices which equal a $K$-dimensional
identity matrix
with its rows permuted and multiplied by either $1$ or $-1$. We find
$\mathbf{T}_b$ by testing all the reorientation matrices and selecting
the one
that minimizes the distance, which for small values of $K$ of interest
in our
application is not computationally expensive. To define our target, we
draw a
``pivot'' $\bolds{\Lambda}_{Vb^{*}}$ at random. We reorient all the
$\bolds{\Lambda}_{Vb}$ to $\bolds{\Lambda}_{Vb^{*}}$. We then
calculate the
vector of mean loadings across all draws under the reorientation
decisions and
use this mean as the pivot in the next iteration of the algorithm. We iterate
until convergence of the mean, which implies convergence of the reorientation
decisions. As a final step, we examine the orientation of the converged
mean and
apply a single sign relabeling step to all draws that gives the varimax loadings
a desired interpretation. We refer to the final reoriented varimax
loadings by
$\{ \bolds{\Lambda}_{Fb} \}$. In Section~\ref{loadings} and in the
supplemental material [\citet{lsm:befa:supplement}], we present
evidence that our
algorithm successfully translated the $\{ \bolds{\Lambda}_{Vb} \}$
into a
common, interpretable orientation for the $\{ \bolds{\Lambda}_{Fb} \}
$. Our
approach is similar to the method of \citet{hoff:raft:hand:2002}. They
also use
equation~(\ref{eq:orient}) to select loadings; however, they use the
criterion to
select not only the column permutations and sign reflections, but also the
rotation. They find a closed form for the solution. Because we want to
use the
varimax rotation, we cannot use their solution. They also use an external
target. Because we do not have such a target, we use our iterative procedure
instead.

Rotation of the working loadings $\{ \bolds{\Lambda}_{b} \}$ to the final
varimax loadings $\{ \bolds{\Lambda}_{Fb} \}$ necessitates rotation
of the
sampled factor scores $\{ \bolds{\eta}_{b} \}$ to factor scores $\{
\bolds{\eta}_{Fb} \}$ concordant with final loadings. Elementary linear
algebra can be used to show that the required orthogonal rotation is
$\bolds{\eta}_{Fb} = \bolds{\Lambda}_{Fb}^{\prime} \bolds{\Lambda}_{b}
(\bolds{\Lambda}_{b}^{\prime} \bolds{\Lambda}_{b})^{-1} \bolds
{\eta}_{b}$. We use
these factor scores in our second stage analysis examining the relationships
between latent teaching constructs inferred from the classroom observation
scores and other teacher quality indicators.

Taken together, our three-step approach (exchangeable prior distributions,
draw-by-draw varimax rotation and reorientation of varimax draws to a common
orientation) provides Bayesian EFA inferences that are invariant to permutations
of the data dimensions. The chosen prior distributions provide
permutation-invariant posterior distributions for $\mathbf{Q}$ and
$\mathbf{U}$.
The varimax criterion is itself permutation invariant because it is constant
across reordering of rows. Finally, the relabeling algorithm depends on only
Euclidean distances and, consequently, behaves identically across different
orders of the variables. Thus, we can be confident that our inferences about
the factor structure, loadings and factor scores are not sensitive to the
arbitrary choice about how the variables are ordered.

\section{Analysis of instructional ratings data}\label{case}

\subsection{Model selection} \label{fit}

Our model assumes a known number of factors $K$, but we need to
determine $K$
from our data. We evaluated possible values of $K$ using the log pseudo marginal
likelihood (LPML) leave-one-out fit statistic as described in
\citet{congdon:2005}. The LPML calculations use importance sampling reweighting
of the posterior distributions over model parameters to estimate the conditional
predictive ordinate $f(\mathbf{y}_{i}|\mathbf{y}_{-i},K)$
[\citet{Geis:Eddy:1979}], where $\mathbf{y}_{-i}$
denotes all data vectors excluding $y_{i}$. The LPML for a given value
of $K$
is then defined as $\log (\prod_{i=1}^{N} f(\mathbf{y}_{i}|\mathbf{y}_{-i},K) )$. The
leave-one-out property induces a penalty for model complexity and helps to
assess the possibility for overfitting.

The LPML statistic has nontrivial Monte Carlo error for chains of the length
that we could feasibly post-process. Hence, we based our calculations
on five
independent chains for each $K=1, \ldots, 5$ and for each subject. We average
values across chains to produce our final LPML estimates for each $K$ and
subject. We adapted each chain for 1000 iterations, and then ran each
chain for
an additional 80,000 iterations, discarding the first 50,000 for
burn-in. We
used the Gelman--Rubin statistics to assess convergence of the elements of
$\mathbf{Q}$ and $\mathbf{U}$ and they all had values near 1. Posterior
sampling for our models is conducted in the \emph{Just Another Gibbs Sampler}
(JAGS) platform of \citet{JAGS}.

To further evaluate the appropriate number of factors, we also examined the
eigenvalues of the correlation matrix for $\bolds{\delta}$. To
estimate the
eigenvalues, we fit the EFA model with $K=10$ factors at the teacher level,
calculated the correlation matrix and its eigenvalues from each draw of
$\mathbf{Q} + \mathbf{U}$, and used the posterior distribution of the ordered
eigenvalues for our inferences.
We used Horn's parallel analysis [\citet{horn:1965}] which compares the
estimated
eigenvalues to those that would be obtained if the dimensions were actually
independent. Let $\tilde{\xi}_1, \ldots, \tilde{\xi}_{10}$ equal
the posterior
means of the ordered eigenvalues of $\mathbf{Q} + \mathbf{U}$. We generated
100,000 independent samples of $N_{\mathrm{teach}}$ $D$-dimensional independent
Gaussian random vectors and for each sample estimated the ordered
eigenvalues of
the sample correlation matrix. Let $\hat{\xi}_1, \ldots, \hat{\xi
}_{10}$ equal
the 95th percentiles across the 100,000 samples of the first 10 ordered
eigenvalues. Horn's parallel analysis selects $K$ as the largest value
such that
$\tilde{\xi}_K > \hat{\xi}_K$, that is, the largest $K$ for which the
corresponding eigenvalue estimated from the data would be unlikely to
occur if
the dimensions were truly independent.
Finally, we also evaluated the simple structure of the loadings for
interpretability, examined their credible intervals, and compared the factor
scores to the teacher knowledge test scores and student achievement
growth to
assess whether the factors appeared to be identifying meaningful
attributes of
teaching.

%
\begin{figure}

\includegraphics{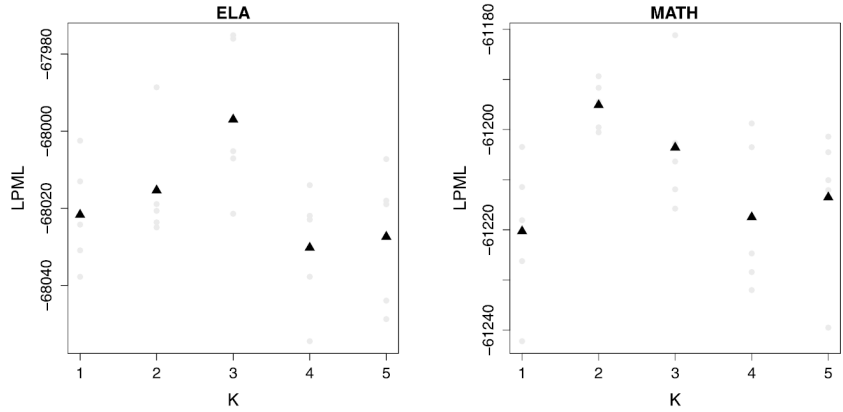}

\caption{Estimated LPML by subject for models with $K=1,\ldots,5$
factors. Black
triangles equal the average from five independent chains and gray dots are
the values for each chain. Larger values indicate better
fit.} \label{fig:lpml}
\end{figure}

%
\begin{figure}[b]

\includegraphics{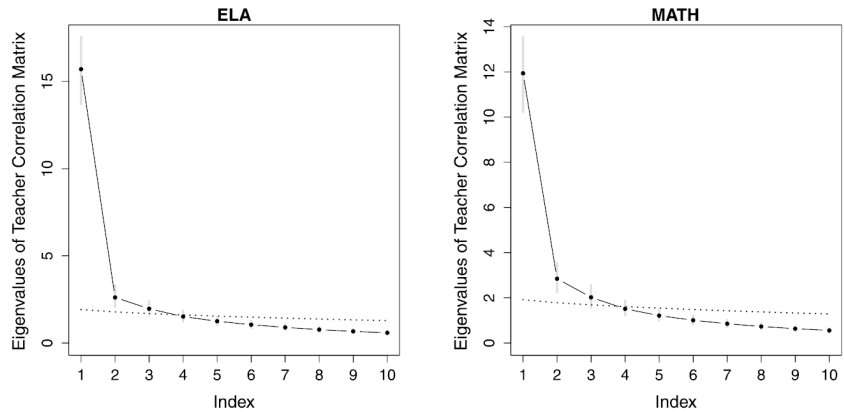}

\caption{Horn parallel analysis to assess the number of factors by
subject. Dots equal the posterior mean of the eigenvalues of the estimated
correlation matrix for latent teacher level dimension scores from a model
with $K=10$. Gray bars are the 95\% credible intervals for the
eigenvalues. The dotted line is the 95th percentile for the eigenvalues
of a
correlation matrix estimated from a sample of $D$-dimensional vectors of
independent random Gaussian variables. The suggested number of factors is
the largest value of $K$ such that the corresponding mean eigenvalue is
greater than the dotted line.}\label{fig:horn}
\end{figure}

Figure~\ref{fig:lpml} presents the estimated LPML for both math and
ELA. Since
larger values of LPML indicate better fit, for both subjects, $K > 3$
is clearly
too many factors. For math, $K=1$ appears to yield a poorly fitting
model as
well. The best fit for math is for $K=2$, but the variability across
chains is
large for $K=3$ and the fit statistic does not rule out $K=3$. Also, as
shown in
Figure~\ref{fig:horn}, the parallel analysis suggests $K=3$ as a plausible
number of factors because the posterior mean of the fourth eigenvalue
is below
the corresponding bound. Hence, we estimate the loadings and compare factor
scores from fits with $K=2$ and $3$. For ELA, $K=3$ yields the largest average
LPML across the five chains, but there is sufficient noise so that
$K=2$ and
perhaps even $K=1$ cannot be ruled out. The parallel analysis again suggests
$K=3$. We thus explore models with $K=1$, $2$ and $3$ and present
results for
$K=2$ and $3$.

\subsection{Identifying constructs of high-quality teaching}\label{loadings}

For each subject and for each of $K=2$ and $3$, we calculated posterior
distributions of reoriented varimax loadings, and corresponding factor scores,
using the procedure given in Section~\ref{ssec:loadings}. We validated
that the
reorientation step was functioning well using three criteria. The first
confirmed that unlike the ``raw'' distributions of varimax solutions (before
reorientation), which were multimodal due to the sign and column indeterminacy,
the reorientation produced unimodal, approximately symmetric
distributions for
the loadings. We used both visual inspection of the densities and the ``dip''
test [\citet{hartigan:1985}] to test for unimodality. The dip test rejected
unimodality for most of the raw varimax distributions, with $p$-values near
zero, but the $p$-values for the tests on the reoriented distributions were
almost all nearly one. Second, we confirmed that the MCMC samples of reoriented
loadings vectors were generally close (in Euclidean distance) to the posterior
mean loading vector, whereas prior to reorientation, the distances of individual
draws to the posterior mean were larger and multimodal, again due to
sign and
column indeterminacy of the raw varimax solutions. Third, we used
multidimensional scaling to confirm that groups of MCMC samples of the raw
varimax solutions that were clustered together in multidimensional space
received the same reorientation decision. These investigations involve
a large
number of plots that are presented in the supplemental material, along with
additional details on the assessment of unimodality of the loadings
distributions [\citet{lsm:befa:supplement}]. Finally, we ran our algorithm
multiple times with different choices for the initial pivot and the inferences
about the loadings were unaffected.

\begin{figure}

\includegraphics{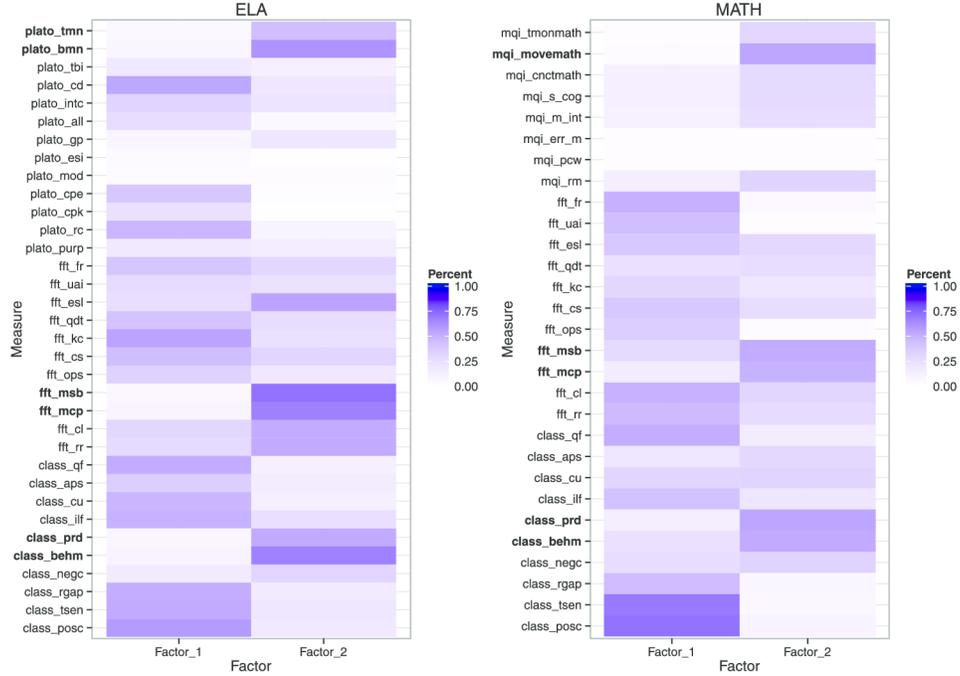}

\caption{Posterior mean varimax loadings normalized to percentage of
variance explained for $K = 2$.}\vspace*{-4pt}\label{fig:tile2}
\end{figure}

\begin{figure}

\includegraphics{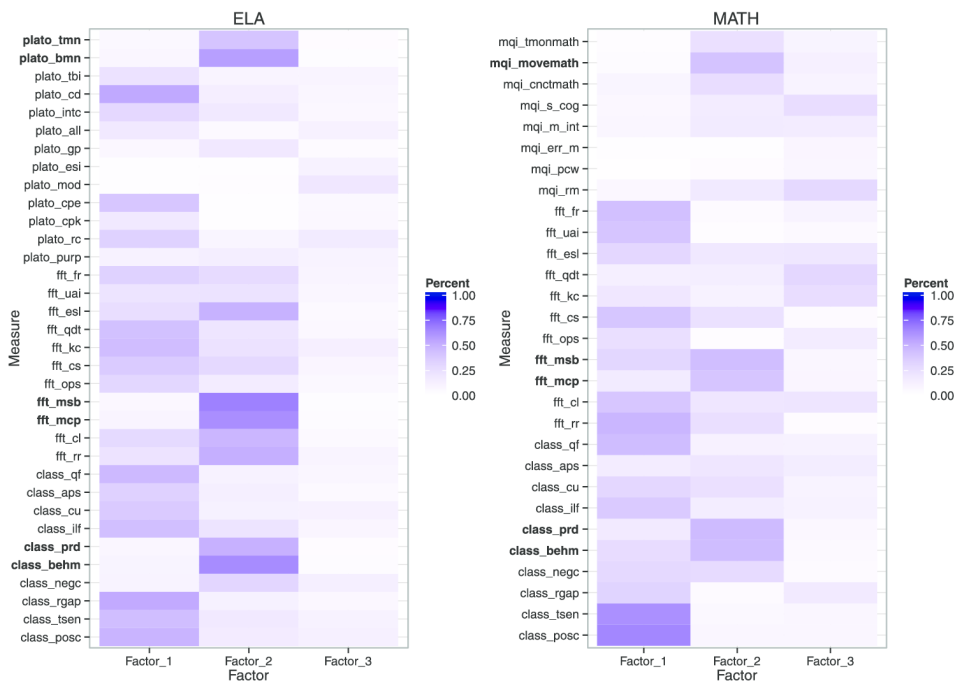}

\caption{Posterior mean varimax loadings normalized to percentage of
variance explained for $K = 3$.}\vspace*{-3pt}\label{fig:tile3}
\end{figure}

The resulting loadings for $K=2$ and $3$ are presented in
Figures~\ref{fig:tile2} and~\ref{fig:tile3}. The figures show the standardized
squared loadings by factor for each dimension of all the protocols.
Dark values
indicate a large loading that explains a large proportion of the
variability in
the latent teacher-level dimension score. Light values indicate little variance
is explained by the factor and a weak loading. For both math and ELA, the
loadings on the third factor when $K=3$ in Figure~\ref{fig:tile3} are generally
weak for all dimensions. For ELA, all of the 95 percent credible
intervals for
the loadings on the third factor include zero (i.e., none of the
loadings are
significant) and for math, only one loading is significant. This is in contrast
to the first two factors, which each have multiple dimensions with clearly
positive loadings.

%

%

Moreover, the loadings patterns for the first two factors for $K=3$ are nearly
identical to those for $K=2$. In both cases, dimensions from all
protocols that
are related to management of student behavior and productivity, in the
sense of
keeping the classroom on task, load heavily on the second factor. These include
the Behavior Management and Productivity dimensions of CLASS, the
Management of
Student Behaviors and Management of Classroom Procedures for FFT, the
MQI Moves
Math Along indicator for math, and the PLATO Time Management and Behavioral
Management dimensions for ELA (the labels of which are bold in the
figures). All
of the protocols assess the teacher's ability to manage the class, and
they are
finding a common attribute that is distinct from the other underlying features
of teaching. Similarly, the dimensions from all protocols that are
related to
instructional quality and student support load heavily on the first factor.
Evidently the constructs of teaching assessed in our classroom observation
ratings are the teacher's \emph{Instructional Practices} and support,
and his or
her \emph{Classroom Management}, where we use the italicized labels to
refer to
these constructs for the remainder. Table~\ref{tab:loadings} in the
\hyperref[app]{Appendix}
presents the posterior mean loadings for $K=2$ along with brief
descriptions of
each dimension.

\subsection{Relationships of factors to other teacher measures} \label
{ssec:stage2}

Understanding how, if at all, the latent instructional constructs
derived from
the lesson ratings relate to other indicators of teaching quality is
critical to
assessing the validity of the constructs. If the estimated constructs
relate in
predictable ways to other measures, we can be more confident in the substantive
interpretations of the constructs based on the loadings patterns and the
conclusion that the constructs capture relevant dimensions of instructional
quality. We thus used two other proposed measures of teaching quality---namely,
teacher knowledge and teacher's students' achievement growth---to
explore the
validity of the teaching constructs derived from the instructional practice
ratings.\looseness=-1

First, each teacher in the study was administered a test of content and
pedagogical content knowledge [\citet{shulman:1987:pct}] specific to their
subject-area specialty (math or ELA), which we refer to as ``Teacher Knowledge
(TK).'' The tests consisted of dichotomously scored items ($30$ for ELA and
$38$ for math) drawn from established teacher knowledge assessments. We
fit a
one-parameter item response theory (IRT) model
[\citet{vanderlinden:hambleton:1996}] to estimate teacher knowledge.
The IRT
estimates correlated above 0.97 with the percentage correct, for both
ELA and
math, and had reliabilities of 0.85 for math and 0.78 for ELA.

Second, we constructed measures of ``Teacher Value-Added (TVA)'' for each
teacher in the study. TVA equals the growth in a teacher's students'
standardized achievement test scores. It is typically estimated by a regression
of student test scores on prior year scores and other student background
variables. Such measures are increasingly being used as part of states' and
districts' formal teacher evaluation systems due to the growing belief
that they at
least partially reflect causal relationships between teacher
instruction and
student learning [\citet{MET:2013}]. To calculate TVA, we used
administrative data
collected from the participating school districts. The data include links
between individual students and their teachers and classrooms, and they include
students' background information and standardized test scores on the state's
accountability test, both for the study school years and multiple prior years.
We estimated TVA using the latent regression methods of
\citet{lockwood:mccaffrey:2014:lr}, which regresses outcome test
scores on
teacher indicator variables, student background characteristics and student
prior test scores while accounting for the measurement error in the
prior test
scores. TVA equals the estimated coefficients on the teacher indicator
variables. The reliability of the estimated TVA equals 0.89 for math
and 0.80
for ELA.


To examine the relationships between TK and TVA and the estimated teaching
constructs from the instructional practice ratings, we used the methods
described in Section~\ref{ssec:loadings} to obtain posterior samples
of the
factor scores $\{ \bolds{\eta}_{Fbj} \}$ for each teacher and each of
$K=2$ and
$K=3$. Let $\{\eta_{Fbj1}\}$ equal the sample of \emph{Instructional
Practices}
factor scores for the 231 math teachers for the $K=2$ model. Let
$\hat{\theta}_j$ equal their estimated TK. For each posterior draw,
we estimated
the sample correlation between $\eta_{Fbj1}$ and $\hat{\theta}_j$ as
$C_{1,\mathrm{TK},b}$. To obtain the correlation on the latent
variable scale,
we use $\tilde{C}_{1,\mathrm{TK},b}=C_{1,\mathrm{TK},b}/\sqrt{r}$,
where $r$ is the
estimated reliability of TK. We use $\{ \tilde{C}_{1,\mathrm{TK},b} \}
$ to
approximate a posterior sample of the disattenuated correlation between
the \emph{Instructional Practices} attribute and teacher knowledge. We
then repeated
this procedure with the remaining factor for math and for both ELA
factors. We
also repeated the analysis for TVA and for the factors from the models with
$K=3$.

%
\begin{figure}

\includegraphics{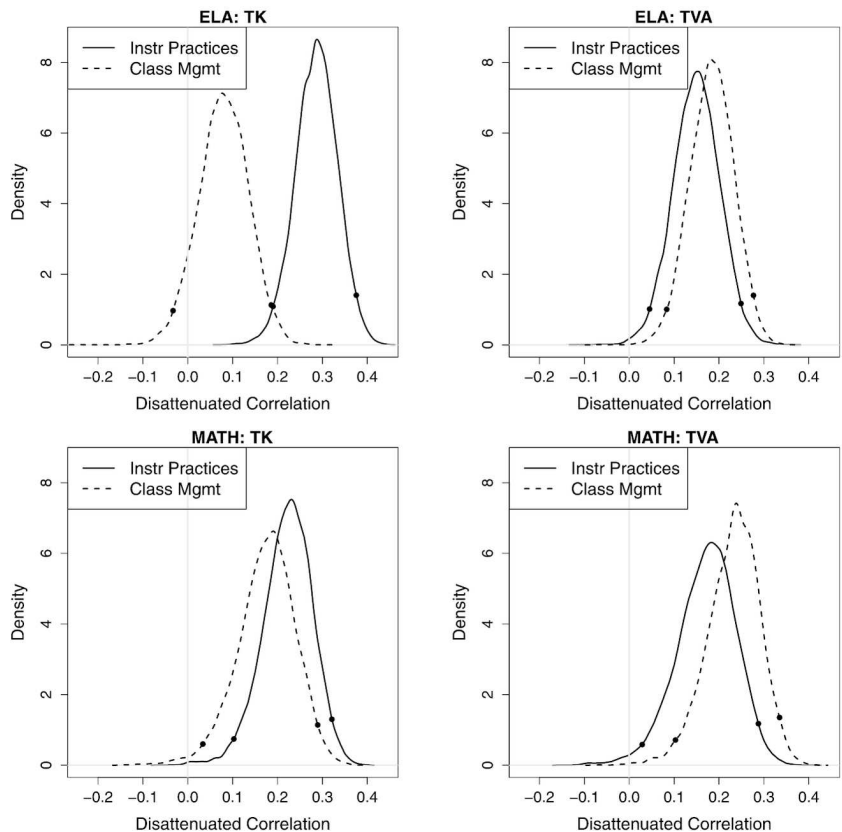}

\caption{Estimated posterior densities of disattenuated correlations between
instructional ratings factors and external measures, by subject (row) and
external measure (column). Different factors given by different line
types within each frame. Dots on the densities correspond to the 0.025
and 0.975 quantiles of each distribution.}\label{fig:stage2}
\end{figure}

Figure~\ref{fig:stage2} plots the estimated posterior densities of these
disattenuated correlations for models with $K=2$. The factor scores for
\emph{Instructional Practices} are related to both TVA and TK, for
both subjects,
with estimated correlations in the 0.15 to 0.30 range. This aligns with
theoretical predictions in that more knowledgeable teachers should be more
capable of providing more effective instruction, which in turn leads to improved
student achievement. The relationships are somewhat stronger with TK
than with
their students' achievement gains. The \emph{Classroom Management}
factor, on the
other hand, is unrelated with TK for ELA teachers, but related to TVA
for both
subjects and to TK in math. The relationship of the \emph{Classroom Management}
factor to TVA is at least as strong as the relationship of \emph{Instructional
Practices} to TVA, and perhaps stronger. The difference between
subjects in
how \emph{Classroom Management} relates to TK may indicate differences
in the
skills necessary to effectively manage math and ELA classes, or it
might reflect
differences in the focus of the observation protocols. For example, the MQI
productivity dimension specifically focuses on keeping the math content moving,
which might require teachers to have sufficient knowledge to retain a
focus on
mathematics. The PLATO dimensions that load on \emph{Classroom
Management} are
very focused on managing behavior and classroom operations and may
require less
content knowledge.

We repeated the analysis using the factor scores from the models with $K=3$.
The inferences for the \emph{Instructional Practices} and \emph{Classroom
Management} factors were virtually identical, consistent with the nearly
identical loadings patterns for these factors in the $K=2$ and $K=3$ models
shown in Figures~\ref{fig:tile2} and~\ref{fig:tile3}. Conversely, the third
factor was not significantly related to either TK or TVA for either subject,
which we interpreted as further evidence that this factor was most likely
spurious.


\section{Discussion}\label{discussion}

We are encouraged that like dimensions across different rating
instruments load
together on the same constructs; for example, the dimensions from different
instruments that connote the management of student behavior all load to
the \emph{Classroom Management} factor in our data. This provides
support for
interpreting the dimensions from different instruments purported to measure
similar constructs as doing so. It also suggests that the instruments
are not
creating spurious differences in the measurement of the primary
constructs of
\emph{Instructional Practices} and \emph{Classroom Management}. This is
practically useful for states and districts having to decide among different
instruments because it suggests that inferences about these broad
domains of
teaching quality may not be very sensitive to the choice.

We are also encouraged that the estimated latent constructs from the
instructional ratings relate in sensible ways to measures of both teacher
knowledge and student achievement outcomes. The \emph{Instructional Practices}
and \emph{Classroom Management} constructs emerge as distinct in the factor
analysis and have some evidence of relating differently to the external
measures. The finding that effective management of student behavior
appears to
be more strongly related to student achievement outcomes than to teacher
knowledge underscores the notion that both effective instruction and effective
behavioral management may be important attributes of classroom
environments that
are successful at promoting student learning.

On the other hand, our results raise some challenging questions given the
significant resource investments being made across the country in
fielding and
using these measures. Our discovery of only two main constructs across
all of
the dimensions that various protocols intend to evaluate raises
questions about
the validity of using scores to differentiate among teachers'
performances on
particular dimensions, an activity valued by stakeholders for targeting
professional development. Perhaps we would discover more constructs
were we to
allow for correlated factors, though the results of \citet{mcca:class:2015}
suggest the correlations among those constructs would be over $0.9$.
Similarly, observing more dimensions might help to differentiate additional
factors. For example, \citet{hamre:2013} hypothesize three domains to classroom
practices: classroom management, emotional support, and instructional support.
The dimensions from the latter two all load onto our \emph
{Instructional Practices}
factor. With additional dimensions specific to each domain we might be
able to
measure them separately. It also may be important for future research
to examine
those dimensions that express relatively large uniqueness variances. Returning
to Figures~\ref{fig:tile2}~and~\ref{fig:tile3}, several dimensions of the
subject-specific protocols (PLATO and MQI) load only weakly on both of our
identified factors and may be capturing important aspects of
instruction that
are particular to their respective subject areas.

Another concern is that while the patterns of correlations of our estimated
factor scores with the other teaching quality indicate help to validate the
constructs, the magnitudes of the correlations are very modest even after
disattenuation for measurement error. For instance, our findings
suggest that
the \emph{Instructional Practices} construct explains less than 10\%
of the
variation among teachers in their effects on student achievement as
measured by
the state's accountability test. Our findings of only modest
correlations among
different modes of measuring teaching quality (e.g., ratings of
instruction and
student achievement outcomes) replicate those of previous studies
[\citet{MET:2013}] and add to a growing body of evidence that there remain
fundamental uncertainties about the constructs that define teaching
quality and
how they can be measured accurately. It is important to stipulate that
it was
not the goal of our analysis to find the combination of dimensions that would
best predict either TVA or TK, but rather to examine whether the factors
determining the communalities of the dimensions behaved sensibly. It is likely
that alternative combinations of the dimensions that included both the
communality and uniqueness of each dimension could lead to better predictions,
although preliminary investigations with our data suggested that the magnitude
of the improvements over the correlations summarized in Figure~\ref{fig:stage2}
are not large.

It is also possible that the modest correlations of the instructional ratings
constructs with other teaching quality indicators may reflect intrinsic
limitations of our observation measures. The dimensions may not fully measure
the practices they intend to evaluate. For example, there may be
infrequent but
high-leverage student--teacher interactions that are critical for enhancing
learning that tend to be missed due to the limited number of
observations on
each teacher. Another example of incomplete measurement is the
evaluation of
classroom management practices, where a high score is ambiguous because
it could
reflect either actively effective management or simply that the
students were
well behaved and the teacher did not have to demonstrate management proficiency.
This ambiguity could be partially responsible for the fact that the dimensions
designed to measure the \emph{Classroom Management} factor tended to have
stronger rater agreement than other dimensions, which in turn could be related
to its emergence as a distinct factor in our analysis. Further
refinements to
the scoring rubrics may improve the ability of the instruments to reliably
distinguish different behaviors.
Finally, the modest correlations of the constructs with student
outcomes as
measured by state standardized exams might also reflect limitations of the
exams. More research is needed to understand to what degree state exams and
student performances on them reflect student learning outcomes that are expected
to be malleable through observable classroom practices.

Our results may also be sensitive to the sample of teachers and schools
participating in the study. The teachers and schools were volunteers.
Given that
teachers knew that their lessons would be observed and rated during the
study, a
potential concern with our sample is that teachers who felt their practices
would not rate highly might have been less likely to participate. Similarly,
principals who were uncertain about their teachers' performances might
have been
more likely to decline our invitation for his/her school to
participate. Such
censoring could attenuate correlations. We do not have classroom practice
measures for all teachers in the participating districts, but we do
have TVA for
all teachers in the districts. The mean TVA for math teachers in our
sample is
about 0.2 standard deviation units greater than the overall mean, and
the mean
TVA for the ELA teachers is about 0.1 standard deviation units greater
than the
overall mean, where standard deviation units are for the latent TVA.
The average
prior achievement in math, reading and language of students in the
participating teachers' classrooms also tended to be higher than the
average for
all the students in the districts. These results are consistent with
the concern
that higher-performing teachers and classes were more likely to
participate. However, the variance of the latent TVA in the sample is
only very
weakly attenuated relative to the variance of the latent TVA for all teachers:
the ratio of the variance for the UTQ teachers to that of all teachers
is 1.0
for ELA teachers and 0.9 for math teachers. Also, \citet
{gitomer:etal:2014} find
that teachers are relatively weak judges of the quality of their classroom
practices, so it is unlikely that teacher self-selection into the study
on the
basis of perceived instructional quality would lead to significant
censoring of
instructional practice ratings. Indeed, our data contain many low
scores on
both instructional practice ratings, as well as on the TK assessments. Our
interpretation is that our sample has sufficient variability to study
relationships among teaching quality measures. Some relationships may be
attenuated, but we suspect any attenuation is not large. Beyond being
volunteers, our study was restricted to middle school math and ELA
teachers in
three large suburban school districts in the same metropolitan area. Conducting
similar studies in other schools, grade levels and subject areas would
help to
understand whether the constructs and relationships we identified
generalize to
other settings.


Our approach to permutation-invariant Bayesian EFA has strengths and weaknesses
for applied research relative to the standard lower triangular
specification. It
is ideally suited to applications where (1) there exists little prior knowledge
for the number and composition of constructs; (2) the amount of data is
modest so
that the potential influence of the prior is a practical concern; and
(3) trying
many different variable orderings is computationally prohibitive. It also
applies to models that do not model factor loadings and scores during
estimation, such as the approach of \citet{carmeci:2008} that directly models
the reduced-rank communality matrix $\mathbf{Q}$. Like the lower triangular
specification, our approach requires few hyperparameter settings, no
tuning of
the sampler, and is readily implemented in standard BUGS language
software. Its
main shortcoming is the need for post hoc identification of the desired
loadings. While post hoc identification is not uncommon, it can lead to
ambiguities in reorientation decisions for individual draws that may hamper
inference when either the sample size is very small or when $K$ is
large. The
lower triangular specification does not have this problem, and
especially when
there are sufficient data to dominate the prior or when the
computational costs
of refitting the model many times are minimal, it may be a more
practical choice
than our method.

Finally, our approach to post hoc reorientation of MCMC draws of working
loadings to achieve simple structure may be of general interest because it
applies not only to our permutation-invariant prior, but also to the lower
triangular specification. It can also be easily adapted to orthogonal rotation
methods other than varimax. Additional work would be required to extend the
approach to oblique rotations, which are often valuable in applications for
improved interpretability of the factors. Also, as noted by
\citet{hahn:carv:scott:2012}, sparsity priors can be beneficial for factor
models, yielding more interpretable loadings and balancing between bias and
variance in exploratory models of structure. For our model, sparsity
can be
obtained by the choice of distribution for components of our loadings
in the
parameter expansion by the methods of \citet{Bhat:Duns:spar:2011} or
\citet{Carv:Pols:Scot:2010}.\vspace*{10pt}

%
\begin{appendix}\label{app}
\section*{Appendix: Posterior mean loadings}\vspace*{-35pt}
%
\begin{table}[b]
\tabcolsep=0pt
\caption{Posterior means of loadings for each subject and dimension
from the $K=2$ models. ``Inst'' denotes \emph{Instructional Practices}
and ``Mgmt'' denotes \emph{Classroom Management}}\label{tab:loadings}
\begin{tabular*}{\textwidth}{@{\extracolsep{\fill}}lld{1.2}d{2.2}d{2.2}d{1.2}@{}}
\hline
& & \multicolumn{2}{c}{\textbf{ELA}} &\multicolumn{2}{c@{}}{\textbf{Math}}\\ [-6pt]
& & \multicolumn{2}{c}{\hrulefill} &\multicolumn{2}{c@{}}{\hrulefill}\\
\textbf{Instrument} & \multicolumn{1}{c}{\textbf{Dimension}} & \multicolumn{1}{c}{\textbf{Inst}} & \multicolumn{1}{c}{\textbf{Mgmt}} & \multicolumn{1}{c}{\textbf
{Inst}} & \multicolumn{1}{c@{}}{\textbf{Mgmt}} \\
\hline
MQI & richness of math content (rm) & & & 0.18 & 0.28
\\
& procedural and computational work (pcw) & & & -0.04 & 0.02 \\
& no errors in mathematics (err\_m) & & & 0.05 & 0.04 \\
& math interactions with students (m\_int) & & & 0.14 & 0.23 \\
& student cognitive demand (s\_cog) & & & 0.17 & 0.27 \\
& class work connected to math (cnctmath) & & & -0.21 & 0.33 \\
& moving the math along (movemath) & & & 0.06 & 0.46 \\
& time spent on math (tmonmath) & & & 0.05 & 0.27 \\ 
\hline
\end{tabular*}
\end{table}
\setcounter{table}{1}
\begin{table}
\tabcolsep=0pt
\caption{(Continued)}
\begin{tabular*}{\textwidth}{@{\extracolsep{\fill}}lld{1.2}d{2.2}d{2.2}d{1.2}@{}}
\hline
& & \multicolumn{2}{c}{\textbf{ELA}} &\multicolumn{2}{c@{}}{\textbf{Math}}\\ [-6pt]
& & \multicolumn{2}{c}{\hrulefill} &\multicolumn{2}{c@{}}{\hrulefill}\\
\textbf{Instrument} & \multicolumn{1}{c}{\textbf{Dimension}} & \multicolumn{1}{c}{\textbf{Inst}} & \multicolumn{1}{c}{\textbf{Mgmt}} & \multicolumn{1}{c}{\textbf
{Inst}} & \multicolumn{1}{c@{}}{\textbf{Mgmt}} \\
\hline
PLATO & demonstrate purpose (purp) & 0.18 & 0.16 & &\\
& representation of content (rc) & 0.36 & 0.15 & & \\
& connections to prior academic knowledge (cpk) & 0.18 & 0.04 & & \\
& connections to prior personal experience (cpe) & 0.33 & 0.08 & & \\
& use of models and modeling (mod) & 0.06 & -0.06 & & \\
& explicit strategy instruction (esi) & 0.10 & 0.03 & & \\
& guided practice (gp) & 0.10 & 0.17 & & \\
& accommodations for language learners (all) & 0.24 & 0.11 & & \\
& intellectual content (intc) & 0.26 & 0.21 & & \\
& classroom discourse (cd) & 0.48 & 0.28 & & \\
& text-based instruction (tbi) & 0.23 & 0.20 & & \\
& behavioral management (bmn) & 0.20 & 0.63 & & \\
& time management (tmn) & 0.12 & 0.35 & & \\[3pt]
FFT & create environment of respect, rapport (rr) &
0.70 & 0.96 & 0.83 & 0.62 \\
& establish a culture of learning (cl) & 0.76 & 0.99 & 0.82 & 0.64 \\
& manage classroom procedures (mcp) & 0.29 & 0.82 & 0.30 & 0.58 \\
& manage student behavior (msb) & 0.32 & 1.15 & 0.75 & 1.04 \\
& organize physical space (ops) & 0.49 & 0.36 & 0.42 & 0.09 \\
& communicate with students (cs) & 0.76 & 0.64 & 0.57 & 0.46 \\
& demonstrate content knowledge (kc) & 0.90 & 0.59 & 0.40 & 0.31 \\
& use question and discussion techniques (qdt) & 0.61 & 0.47 & 0.33 &
0.35 \\
& engage students in learning (esl) & 0.55 & 0.83 & 0.67 & 0.57 \\
& use assessment in instruction (uai) & 0.39 & 0.35 & 0.54 & 0.08 \\
& flexibility and responsiveness (fr) & 0.64 & 0.55 & 0.61 & 0.17 \\[3pt]
CLASS & positive climate (posc) & 0.67 & 0.36 & 0.76
& 0.24 \\
& teacher sensitivity (tsen) & 0.47 & 0.29 & 0.54 & 0.13 \\
& regard for adolescent perspective (rgap) & 0.43 & 0.24 & 0.34 & 0.11
\\
& negative climate (negc) & 0.30 & 0.43 & 0.38 & 0.43 \\
& behavior management (behm) & 0.25 & 0.65 & 0.38 & 0.59 \\
& productivity (prd) & 0.15 & 0.40 & 0.20 & 0.43 \\
& instructional learning formats (ilf) & 0.44 & 0.31 & 0.32 & 0.22 \\
& content understanding (cu) & 0.36 & 0.18 & 0.29 & 0.29 \\
& analysis and problem solving (aps) & 0.30 & 0.20 & 0.22 & 0.27 \\
& quality of feedback (qf) & 0.39 & 0.19 & 0.38 & 0.20 \\
\hline
\end{tabular*}
\end{table}
\end{appendix}
\newpage

\section*{Acknowledgments}
The authors thank the Associate Editor and two
anonymous reviewers for helpful
comments on earlier drafts.

\begin{supplement}[id=suppA]
\stitle{Supplement to ``Inferring constructs of effective teaching
from classroom observations: {A}n application of {B}ayesian exploratory
factor analysis without restrictions''}
\slink[doi]{10.1214/15-AOAS833SUPP} 
\sdatatype{.pdf}
\sfilename{aoas833\_supp.pdf}
\sdescription{This document contains detailed evidence on the
effectiveness of our reorientation algorithm for the
varimax loadings.}
\end{supplement}

%

\printaddresses

\begin{thebibliography}{40}

\bibitem[\protect\citeauthoryear{Albert and Chib}{1993}]{Albe:Chib:1993}
%
\begin{barticle}[mr]
\bauthor{\bsnm{Albert},~\bfnm{James~H.}\binits{J.~H.}} \AND
\bauthor{\bsnm{Chib},~\bfnm{Siddhartha}\binits{S.}}
(\byear{1993}).
\btitle{Bayesian analysis of binary and polychotomous response data}.
\bjournal{J. Amer. Statist. Assoc.}
\bvolume{88}
\bpages{669--679}.
\bid{issn={0162-1459}, mr={1224394}}
\end{barticle}
%
\bptok{imsref}%
\endbibitem

\bibitem[\protect\citeauthoryear{Bhattacharya and
Dunson}{2011}]{Bhat:Duns:spar:2011}
%
\begin{barticle}[mr]
\bauthor{\bsnm{Bhattacharya},~\bfnm{A.}\binits{A.}} \AND
\bauthor{\bsnm{Dunson},~\bfnm{D.~B.}\binits{D.~B.}}
(\byear{2011}).
\btitle{Sparse {B}ayesian infinite factor models}.
\bjournal{Biometrika}
\bvolume{98}
\bpages{291--306}.
\bid{doi={10.1093/biomet/asr013}, issn={0006-3444}, mr={2806429}}
\end{barticle}
%
\bptok{imsref}%
\endbibitem

\bibitem[\protect\citeauthoryear{Bill and Melinda Gates
Foundation}{2013}]{MET:2013}
%
\begin{bmisc}[author]
\borganization{Bill and Melinda Gates Foundation}
(\byear{2013}).
\bhowpublished{Ensuring fair and reliable measures of effective
teaching: Culminating findings from the MET project's three-year study.
Available at \url{http://www.metproject.org}.}
\end{bmisc}
%
\bptok{imsref}%
\endbibitem

\bibitem[\protect\citeauthoryear{Browne}{2001}]{browne:2001}
%
\begin{barticle}[author]
\bauthor{\bsnm{Browne},~\bfnm{M.~W.}\binits{M.~W.}}
(\byear{2001}).
\btitle{An overview of analytic rotation in exploratory factor analysis}.
\bjournal{Multivariate Behavioral Research}
\bvolume{36}
\bpages{111--150}.
\end{barticle}
%
\bptok{imsref}%
\endbibitem

\bibitem[\protect\citeauthoryear{Carmeci}{2009}]{carmeci:2008}
%
\begin{bmisc}[author]
\bauthor{\bsnm{Carmeci},~\bfnm{G.}\binits{G.}}
(\byear{2009}).
\bhowpublished{A Metropolis--Hastings algorithm for reduced
rank covariance
matrices with application to Bayesian factor models.
DISES working papers, Univ. Trieste, Italy.}
\end{bmisc}
%
\bptok{imsref}%
\endbibitem

\bibitem[\protect\citeauthoryear{Carvalho, Polson and
Scott}{2010}]{Carv:Pols:Scot:2010}
%
\begin{barticle}[mr]
\bauthor{\bsnm{Carvalho},~\bfnm{Carlos~M.}\binits{C.~M.}},
\bauthor{\bsnm{Polson},~\bfnm{Nicholas~G.}\binits{N.~G.}} \AND
\bauthor{\bsnm{Scott},~\bfnm{James~G.}\binits{J.~G.}}
(\byear{2010}).
\btitle{The horseshoe estimator for sparse signals}.
\bjournal{Biometrika}
\bvolume{97}
\bpages{465--480}.
\bid{doi={10.1093/biomet/asq017}, issn={0006-3444}, mr={2650751}}
\end{barticle}
%
\bptok{imsref}%
\endbibitem

\bibitem[\protect\citeauthoryear{Carvalho
et~al.}{2008}]{Carv:Chan:Luca:Nevi:Wang:West:high:2008}
%
\begin{barticle}[mr]
\bauthor{\bsnm{Carvalho},~\bfnm{Carlos~M.}\binits{C.~M.}},
\bauthor{\bsnm{Chang},~\bfnm{Jeffrey}\binits{J.}},
\bauthor{\bsnm{Lucas},~\bfnm{Joseph~E.}\binits{J.~E.}},
\bauthor{\bsnm{Nevins},~\bfnm{Joseph~R.}\binits{J.~R.}},
\bauthor{\bsnm{Wang},~\bfnm{Quanli}\binits{Q.}} \AND
\bauthor{\bsnm{West},~\bfnm{Mike}\binits{M.}}
(\byear{2008}).
\btitle{High-dimensional sparse factor modeling: Applications in gene
expression genomics}.
\bjournal{J. Amer. Statist. Assoc.}
\bvolume{103}
\bpages{1438--1456}.
\bid{doi={10.1198/016214508000000869}, issn={0162-1459}, mr={2655722}}
\end{barticle}
%
\bptok{imsref}%
\endbibitem

\bibitem[\protect\citeauthoryear{Casabianca, Lockwood and
McCaffrey}{2015}]{casa:lock:mcca:2015}
%
\begin{barticle}[author]
\bauthor{\bsnm{Casabianca},~\bfnm{J.}\binits{J.}},
\bauthor{\bsnm{Lockwood},~\bfnm{J.~R.}\binits{J.~R.}} \AND
\bauthor{\bsnm{McCaffrey},~\bfnm{D.~F.}\binits{D.~F.}}
(\byear{2015}).
\btitle{Trends in classroom observation scores}.
\bjournal{Educational and Psychological Measurement}
\bvolume{75}
\bpages{311--337}.
\end{barticle}
%
\bptok{imsref}%
\endbibitem

\bibitem[\protect\citeauthoryear{Congdon}{2005}]{congdon:2005}
%
\begin{bbook}[mr]
\bauthor{\bsnm{Congdon},~\bfnm{Peter}\binits{P.}}
(\byear{2005}).
\btitle{Bayesian Models for Categorical Data}.
\bpublisher{Wiley},
\blocation{Chichester}.
\bid{doi={10.1002/0470092394}, mr={2191351}}
\end{bbook}
%
\bptok{imsref}%
\endbibitem

\bibitem[\protect\citeauthoryear{Danielson}{2011}]{fft}
%
\begin{bbook}[author]
\bauthor{\bsnm{Danielson},~\bfnm{C.}\binits{C.}}
(\byear{2011}).
\btitle{{Enhancing Professional Practice: A Framework for Teaching}}.
\bpublisher{ASCD},
\blocation{Alexandria, VA}.
\end{bbook}
%
\bptok{imsref}%
\endbibitem

\bibitem[\protect\citeauthoryear{Erosheva and
Curtis}{2013}]{erosheva:curtis:2013}
%
\begin{btechreport}[author]
\bauthor{\bsnm{Erosheva},~\bfnm{Elena~A.}\binits{E.~A.}} \AND
\bauthor{\bsnm{Curtis},~\bfnm{S.~McKay}\binits{S.~M.}}
(\byear{2013}).
\btitle{{Dealing with rotational invariance in Bayesian
confirmatory factor models}}.
\btype{Technical Report} \bnumber{589},
\bnote{Univ. Washington, Seattle, WA}.
\end{btechreport}
%
\bptok{imsref}%
\endbibitem

\bibitem[\protect\citeauthoryear{Fr{\"{u}}wirth-Schnatter and
Lopes}{2013}]{fsl:2013}
%
\begin{bmisc}[author]
\bauthor{\bsnm{Fr\"uwirth-Schnatter},~\bfnm{Sylvia}\binits{S.}}
\AND
\bauthor{\bsnm{Lopes},~\bfnm{Hedibert~Freitas}\binits{H.~F.}}
(\byear{2013}).
\bhowpublished{Parsimonious {B}ayesian factor analysis when the number
of factors is unknown.
Working paper, Univ. Chicago Booth School of Business, Chicago, IL}.
\end{bmisc}
%
\bptok{imsref}%
\endbibitem

\bibitem[\protect\citeauthoryear{Geisser and Eddy}{1979}]{Geis:Eddy:1979}
%
\begin{barticle}[mr]
\bauthor{\bsnm{Geisser},~\bfnm{Seymour}\binits{S.}} \AND
\bauthor{\bsnm{Eddy},~\bfnm{William~F.}\binits{W.~F.}}
(\byear{1979}).
\btitle{A predictive approach to model selection}.
\bjournal{J. Amer. Statist. Assoc.}
\bvolume{74}
\bpages{153--160}.
\bid{issn={0003-1291}, mr={0529531}}
\end{barticle}
%
\bptok{imsref}%
\endbibitem

\bibitem[\protect\citeauthoryear{Gelman}{2006}]{Gelm:2006}
%
\begin{barticle}[mr]
\bauthor{\bsnm{Gelman},~\bfnm{Andrew}\binits{A.}}
(\byear{2006}).
\btitle{Prior distributions for variance parameters in hierarchical
models (comment on article by {B}rowne and {D}raper)}.
\bjournal{Bayesian Anal.}
\bvolume{1}
\bpages{515--533 (electronic)}.
\bid{mr={2221284}}
\bptnote{check related, check pages}%
\end{barticle}
%
\bptok{imsref}%
\endbibitem

\bibitem[\protect\citeauthoryear{Geweke and Zhou}{1996}]{Gewe:Zhou:meas:1996}
%
\begin{barticle}[author]
\bauthor{\bsnm{Geweke},~\bfnm{J.}\binits{J.}} \AND
\bauthor{\bsnm{Zhou},~\bfnm{Guofu}\binits{G.}}
(\byear{1996}).
\btitle{Measuring the pricing error of the arbitrage pricing theory}.
\bjournal{The Review of Financial Studies}
\bvolume{9}
\bpages{557--587}.
\end{barticle}
%
\bptok{imsref}%
\endbibitem

\bibitem[\protect\citeauthoryear{Ghosh and
Dunson}{2009}]{Ghos:Duns:defa:2009}
%
\begin{barticle}[mr]
\bauthor{\bsnm{Ghosh},~\bfnm{Joyee}\binits{J.}} \AND
\bauthor{\bsnm{Dunson},~\bfnm{David~B.}\binits{D.~B.}}
(\byear{2009}).
\btitle{Default prior distributions and efficient posterior
computation in {B}ayesian factor analysis}.
\bjournal{J. Comput. Graph. Statist.}
\bvolume{18}
\bpages{306--320}.
\bid{doi={10.1198/jcgs.2009.07145}, issn={1061-8600}, mr={2749834}}
\end{barticle}
%
\bptok{imsref}%
\endbibitem

\bibitem[\protect\citeauthoryear{Gitomer et~al.}{2014}]{gitomer:etal:2014}
%
\begin{barticle}[author]
\bauthor{\bsnm{Gitomer},~\bfnm{D.~H.}\binits{D.~H.}},
\bauthor{\bsnm{Bell},~\bfnm{C.~A.}\binits{C.~A.}},
\bauthor{\bsnm{Qi},~\bfnm{Y.}\binits{Y.}},
\bauthor{\bsnm{McCaffrey},~\bfnm{D.~F.}\binits{D.~F.}},
\bauthor{\bsnm{Hamre},~\bfnm{B.~K.}\binits{B.~K.}} \AND
\bauthor{\bsnm{Pianta},~\bfnm{R.~C.}\binits{R.~C.}}
(\byear{2014}).
\btitle{The instructional challenge in improving teaching quality:
{L}essons from a classroom observation protocol}.
\bjournal{Teachers College Record}
\bvolume{116}
\bpages{1--32}.
\end{barticle}
%
\bptok{imsref}%
\endbibitem

\bibitem[\protect\citeauthoryear{Gordon, Kane and
Staiger}{2006}]{Gord:Kane:Stai:2006}
%
\begin{bmisc}[author]
\bauthor{\bsnm{Gordon},~\bfnm{R.}\binits{R.}},
\bauthor{\bsnm{Kane},~\bfnm{T.~J.}\binits{T.~J.}} \AND
\bauthor{\bsnm{Staiger},~\bfnm{D.~O.}\binits{D.~O.}}
(\byear{2006}).
\bhowpublished{Identifying effective teachers using performance on the job.
Discussion Paper 2006-01, The Brookings Institution, Washington, DC}.
\end{bmisc}
%
\bptok{imsref}%
\endbibitem

\bibitem[\protect\citeauthoryear{Grossman et~al.}{2010}]{grossman:2010}
%
\begin{bmisc}[author]
\bauthor{\bsnm{Grossman},~\bfnm{P.}\binits{P.}},
\bauthor{\bsnm{Loeb},~\bfnm{S.}\binits{S.}},
\bauthor{\bsnm{Cohen},~\bfnm{J.}\binits{J.}},
\bauthor{\bsnm{Hammerness},~\bfnm{K.}\binits{K.}},
\bauthor{\bsnm{Wyckoff},~\bfnm{J.}\binits{J.}},
\bauthor{\bsnm{Boyd},~\bfnm{D.}\binits{D.}} \AND
\bauthor{\bsnm{Lankford},~\bfnm{H.}\binits{H.}}
(\byear{2010}).
\bhowpublished{Measure for measure: The relationship between measures
of instructional practice in middle school English language arts and
teachers' value-added scores.
Working Paper 16015, National Bureau of Economic Research, Cambridge, MA}.
\end{bmisc}
%
\bptok{imsref}%
\endbibitem

\bibitem[\protect\citeauthoryear{Hahn, Carvalho and
Scott}{2012}]{hahn:carv:scott:2012}
%
\begin{barticle}[mr]
\bauthor{\bsnm{Hahn},~\bfnm{P.~Richard}\binits{P.~R.}},
\bauthor{\bsnm{Carvalho},~\bfnm{Carlos~M.}\binits{C.~M.}} \AND
\bauthor{\bsnm{Scott},~\bfnm{James~G.}\binits{J.~G.}}
(\byear{2012}).
\btitle{A sparse factor analytic probit model for congressional voting
patterns}.
\bjournal{J. R. Stat. Soc. Ser. C. Appl. Stat.}
\bvolume{61}
\bpages{619--635}.
\bid{doi={10.1111/j.1467-9876.2012.01044.x}, issn={0035-9254}, mr={2960741}}
\end{barticle}
%
\bptok{imsref}%
\endbibitem\

\bibitem[\protect\citeauthoryear{Hamre et~al.}{2012}]{hamre:2012}
%
\begin{barticle}[author]
\bauthor{\bsnm{Hamre},~\bfnm{B.~K.}\binits{B.~K.}},
\bauthor{\bsnm{Pianta},~\bfnm{R.~C.}\binits{R.~C.}},
\bauthor{\bsnm{Burchinal},~\bfnm{M.}\binits{M.}},
\bauthor{\bsnm{Field},~\bfnm{S.}\binits{S.}},
\bauthor{\bsnm{LoCasale-Crouch},~\bfnm{J.}\binits{J.}},
\bauthor{\bsnm{Downer},~\bfnm{J.~T.}\binits{J.~T.}},
\bauthor{\bsnm{Howes},~\bfnm{C.}\binits{C.}},
\bauthor{\bsnm{LoParo},~\bfnm{K.}\binits{K.}} \AND
\bauthor{\bsnm{Scott-Little},~\bfnm{C.}\binits{C.}}
(\byear{2012}).
\btitle{A course on effective teacher--child interactions: {E}ffects on
teacher beliefs, knowledge, and observed practice}.
\bjournal{American Educational Research Journal}
\bvolume{49}
\bpages{88--123}.
\end{barticle}
%
\bptok{imsref}%
\endbibitem

\bibitem[\protect\citeauthoryear{Hamre et~al.}{2013}]{hamre:2013}
%
\begin{barticle}[author]
\bauthor{\bsnm{Hamre},~\bfnm{B.~K.}\binits{B.~K.}},
\bauthor{\bsnm{Pianta},~\bfnm{R.~C.}\binits{R.~C.}},
\bauthor{\bsnm{Downer},~\bfnm{J.~T.}\binits{J.~T.}},
\bauthor{\bsnm{DeCoster},~\bfnm{J.}\binits{J.}},
\bauthor{\bsnm{Mashburn},~\bfnm{A.~J.}\binits{A.~J.}},
\bauthor{\bsnm{Jones},~\bfnm{S.~M.}\binits{S.~M.}},
\bauthor{\bsnm{Brown},~\bfnm{J.~L.}\binits{J.~L.}},
\bauthor{\bsnm{Cappella},~\bfnm{E.}\binits{E.}},
\bauthor{\bsnm{Atkins},~\bfnm{M.}\binits{M.}},
\bauthor{\bsnm{Rivers},~\bfnm{S.~E.}\binits{S.~E.}},
\bauthor{\bsnm{Brackett},~\bfnm{M.}\binits{M.}} \AND
\bauthor{\bsnm{Hakigami},~\bfnm{A.}\binits{A.}}
(\byear{2013}).
\btitle{Teaching through interactions: {T}esting a developmental
framework of teacher effectiveness in over 4000 classrooms}.
\bjournal{The Elementary School Journal}
\bvolume{113}
\bpages{461--487}.
\end{barticle}
%
\bptok{imsref}%
\endbibitem

\bibitem[\protect\citeauthoryear{Hartigan and Hartigan}{1985}]{hartigan:1985}
%
\begin{barticle}[mr]
\bauthor{\bsnm{Hartigan},~\bfnm{J.~A.}\binits{J.~A.}} \AND
\bauthor{\bsnm{Hartigan},~\bfnm{P.~M.}\binits{P.~M.}}
(\byear{1985}).
\btitle{The dip test of unimodality}.
\bjournal{Ann. Statist.}
\bvolume{13}
\bpages{70--84}.
\bid{doi={10.1214/aos/1176346577}, issn={0090-5364}, mr={0773153}}
\end{barticle}
%
\bptok{imsref}%
\endbibitem

\bibitem[\protect\citeauthoryear{Hoff, Raftery and
Handcock}{2002}]{hoff:raft:hand:2002}
%
\begin{barticle}[mr]
\bauthor{\bsnm{Hoff},~\bfnm{Peter~D.}\binits{P.~D.}},
\bauthor{\bsnm{Raftery},~\bfnm{Adrian~E.}\binits{A.~E.}} \AND
\bauthor{\bsnm{Handcock},~\bfnm{Mark~S.}\binits{M.~S.}}
(\byear{2002}).
\btitle{Latent space approaches to social network analysis}.
\bjournal{J. Amer. Statist. Assoc.}
\bvolume{97}
\bpages{1090--1098}.
\bid{doi={10.1198/016214502388618906}, issn={0162-1459}, mr={1951262}}
\end{barticle}
%
\bptok{imsref}%
\endbibitem

\bibitem[\protect\citeauthoryear{Horn}{1965}]{horn:1965}
%
\begin{barticle}[pbm]
\bauthor{\bsnm{Horn},~\bfnm{J.~L.}\binits{J.~L.}}
(\byear{1965}).
\btitle{A rationale and test for the number of factors in factor analysis}.
\bjournal{Psychometrika}
\bvolume{30}
\bpages{179--185}.
\bid{issn={0033-3123}, pmid={14306381}}
\end{barticle}
%
\bptok{imsref}%
\endbibitem

\bibitem[\protect\citeauthoryear{Ishwaran}{2000}]{Ishw:univ:2000}
%
\begin{barticle}[mr]
\bauthor{\bsnm{Ishwaran},~\bfnm{Hemant}\binits{H.}}
(\byear{2000}).
\btitle{Univariate and multirater ordinal cumulative link regression
with covariate specific cutpoints}.
\bjournal{Canad. J. Statist.}
\bvolume{28}
\bpages{715--730}.
\bid{doi={10.2307/3315912}, issn={0319-5724}, mr={1821430}}
\end{barticle}
%
\bptok{imsref}%
\endbibitem

\bibitem[\protect\citeauthoryear{Johnson}{1996}]{Johnson:1996}
%
\begin{barticle}[author]
\bauthor{\bsnm{Johnson},~\bfnm{Valen~E.}\binits{V.~E.}}
(\byear{1996}).
\btitle{On {B}ayesian analysis of multirater ordinal data: {A}n
application to automated essay grading}.
\bjournal{J. Amer. Statist. Assoc.}
\bvolume{91}
\bpages{42--51}.
\end{barticle}
%
\bptok{imsref}%
\endbibitem

\bibitem[\protect\citeauthoryear{Kaiser}{1958}]{kaiser:1958}
%
\begin{barticle}[author]
\bauthor{\bsnm{Kaiser},~\bfnm{H.~F.}\binits{H.~F.}}
(\byear{1958}).
\btitle{The varimax criterion for analytic rotation in factor analysis}.
\bjournal{Psychometrika}
\bvolume{23}
\bpages{187--200}.
\end{barticle}
%
\bptok{imsref}%
\endbibitem

\bibitem[\protect\citeauthoryear{{Learning Mathematics for Teaching
Project}}{2006}]{mqi}
%
\begin{bmisc}[author]
\borganization{Learning Mathematics for Teaching Project}
(\byear{2006}).
\bhowpublished{A coding rubric for measuring the mathematics quality
of instruction.
Technical Report LMT1.06, Univ. Michigan, Ann Arbor, MI}.
\end{bmisc}
%
\bptok{imsref}%
\endbibitem

\bibitem[\protect\citeauthoryear{Lockwood and
McCaffrey}{2014}]{lockwood:mccaffrey:2014:lr}
%
\begin{barticle}[author]
\bauthor{\bsnm{Lockwood},~\bfnm{J.~R.}\binits{J.~R.}} \AND
\bauthor{\bsnm{McCaffrey},~\bfnm{D.~F.}\binits{D.~F.}}
(\byear{2014}).
\btitle{Correcting for test score measurement error in ANCOVA models
for estimating treatment effects}.
\bjournal{Journal of Educational and Behavioral Statistics}
\bvolume{39}
\bpages{22--52}.
\end{barticle}
%
\bptok{imsref}%
\endbibitem

\bibitem[\protect\citeauthoryear{Lockwood, Savitsky and
McCaffrey}{2015}]{lsm:befa:supplement}
%
\begin{bmisc}[author]
\bauthor{\bsnm{Lockwood},~\binits{J.}},
\bauthor{\bsnm{Savitsky},~\binits{T.}} \AND
\bauthor{\bsnm{McCaffrey},~\binits{D.}}
(\byear{2015}).
\bhowpublished{Supplement to ``Inferring constructs of effective
teaching from classroom observations:
An application of Bayesian exploratory factor analysis without restrictions.''
DOI:\doiurl{10.1214/15-AOAS833SUPP}}.
\bptok{imsref}%
\end{bmisc}
%
\endbibitem

\bibitem[\protect\citeauthoryear{Lopes and West}{2004}]{Lope:West:baye:2004}
%
\begin{barticle}[mr]
\bauthor{\bsnm{Lopes},~\bfnm{Hedibert~Freitas}\binits{H.~F.}} \AND
\bauthor{\bsnm{West},~\bfnm{Mike}\binits{M.}}
(\byear{2004}).
\btitle{Bayesian model assessment in factor analysis}.
\bjournal{Statist. Sinica}
\bvolume{14}
\bpages{41--67}.
\bid{issn={1017-0405}, mr={2036762}}
\end{barticle}
%
\bptok{imsref}%
\endbibitem

\bibitem[\protect\citeauthoryear{McCaffrey et~al.}{2015}]{mcca:class:2015}
%
\begin{barticle}[author]
\bauthor{\bsnm{McCaffrey},~\bfnm{D.~F.}\binits{D.~F.}},
\bauthor{\bsnm{Yuan},~\bfnm{K.}\binits{K.}},
\bauthor{\bsnm{Savitsky},~\bfnm{T.~D.}\binits{T.~D.}},
\bauthor{\bsnm{Lockwood},~\bfnm{J.~R.}\binits{J.~R.}} \AND
\bauthor{\bsnm{Edelen},~\bfnm{M.~O.}\binits{M.~O.}}
(\byear{2015}).
\btitle{Uncovering multivariate structure in classroom
observations in the presence of rater errors}.
\bjournal{Educational Measurement: Issues and Practice}
\bvolume{34}
\bpages{34--46}.
\end{barticle}
%
\bptok{imsref}%
\endbibitem

\bibitem[\protect\citeauthoryear{McParland et~al.}{2014}]{mcpa:ETAL:2014}
%
\begin{barticle}[mr]
\bauthor{\bsnm{McParland},~\bfnm{Damien}\binits{D.}},
\bauthor{\bsnm{Gormley},~\bfnm{Isobel~Claire}\binits{I.~C.}},
\bauthor{\bsnm{McCormick},~\bfnm{Tyler~H.}\binits{T.~H.}},
\bauthor{\bsnm{Clark},~\bfnm{Samuel~J.}\binits{S.~J.}},
\bauthor{\bsnm{Kabudula},~\bfnm{Chodziwadziwa~Whiteson}\binits
{C.~W.}} \AND
\bauthor{\bsnm{Collinson},~\bfnm{Mark~A.}\binits{M.~A.}}
(\byear{2014}).
\btitle{Clustering {S}outh {A}frican households based on their asset
status using latent variable models}.
\bjournal{Ann. Appl. Stat.}
\bvolume{8}
\bpages{747--776}.
\bid{doi={10.1214/14-AOAS726}, issn={1932-6157}, mr={3262533}}
\end{barticle}
%
\bptok{imsref}%
\endbibitem

\bibitem[\protect\citeauthoryear{Peterson
et~al.}{2011}]{peterson2011globally}
%
\begin{bmisc}[author]
\bauthor{\bsnm{Peterson},~\bfnm{Paul~E.}\binits{P.~E.}},
\bauthor{\bsnm{Woessmann},~\bfnm{Ludger}\binits{L.}},
\bauthor{\bsnm{Hanushek},~\bfnm{Eric~A.}\binits{E.~A.}} \AND
\bauthor{\bsnm{Lastra-Anad\'on},~\bfnm{Carlos~X.}\binits{C.~X.}}
(\byear{2011}).
\bhowpublished{Globally challenged: Are US students ready to compete?
PEPG Report 11-03, Harvard's Program on Education Policy and Governance
\& Education Next,
Taubman Center for State and Local Government, Harvard Kennedy School,
Cambridge, MA}.
\end{bmisc}
%
\bptok{imsref}%
\endbibitem

\bibitem[\protect\citeauthoryear{Plummer}{2003}]{JAGS}
%
\begin{binproceedings}[author]
\bauthor{\bsnm{Plummer},~\bfnm{M.}\binits{M.}}
(\byear{2003}).
\btitle{JAGS: A program for analysis of Bayesian graphical models
using Gibbs sampling}.
In \bbooktitle{Proceedings of the 3rd International Workshop on
Distributed Statistical Computing (DSC 2003)}.
\blocation{Vienna, Austria}.
\end{binproceedings}
%
\bptok{imsref}%
\endbibitem

\bibitem[\protect\citeauthoryear{Savitsky and
McCaffrey}{2014}]{savitsky:mccaffrey:2014}
%
\begin{barticle}[mr]
\bauthor{\bsnm{Savitsky},~\bfnm{Terrance~D.}\binits{T.~D.}} \AND
\bauthor{\bsnm{McCaffrey},~\bfnm{Daniel~F.}\binits{D.~F.}}
(\byear{2014}).
\btitle{Bayesian hierarchical multivariate formulation with factor
analysis for nested ordinal data}.
\bjournal{Psychometrika}
\bvolume{79}
\bpages{275--302}.
\bid{doi={10.1007/s11336-013-9339-z}, issn={0033-3123}, mr={3255120}}
\end{barticle}
%
\bptok{imsref}%
\endbibitem

\bibitem[\protect\citeauthoryear{Shulman}{1987}]{shulman:1987:pct}
%
\begin{barticle}[author]
\bauthor{\bsnm{Shulman},~\bfnm{L.~S.}\binits{L.~S.}}
(\byear{1987}).
\btitle{Knowledge and teaching: {F}oundations of the new reform}.
\bjournal{Harvard Educational Review}
\bvolume{57}
\bpages{1--23}.
\end{barticle}
%
\bptok{imsref}%
\endbibitem

\bibitem[\protect\citeauthoryear{Stephens}{2000}]{stephens:2000}
%
\begin{barticle}[mr]
\bauthor{\bsnm{Stephens},~\bfnm{Matthew}\binits{M.}}
(\byear{2000}).
\btitle{Dealing with label switching in mixture models}.
\bjournal{J. R. Stat. Soc. Ser. B. Stat. Methodol.}
\bvolume{62}
\bpages{795--809}.
\bid{doi={10.1111/1467-9868.00265}, issn={1369-7412}, mr={1796293}}
\end{barticle}
%
\bptok{imsref}%
\endbibitem\

\bibitem[\protect\citeauthoryear{van der Linden and
Hambleton}{1997}]{vanderlinden:hambleton:1996}
%
\begin{bbook}[mr]
\bauthor{\bsnm{{van der Linden}},~\bfnm{W.}\binits{W.}} \AND
\bauthor{\bsnm{Hambleton},~\bfnm{R.~K.}\binits{R.~K.}}, eds.
(\byear{1997}).
\btitle{Handbook of Modern Item Response Theory}.
\bpublisher{Springer},
\blocation{New York}.
\bid{doi={10.1007/978-1-4757-2691-6}, mr={1601043}}
\bptnote{check year}%
\end{bbook}
%
\bptok{imsref}%
\endbibitem
\end{thebibliography}
\end{document}